\documentclass[aps, physrev, reprint, superscriptaddress]{revtex4-2}

\usepackage[utf8]{inputenc}
\usepackage{amsmath, amssymb}
\usepackage{graphicx}
\usepackage{xcolor}
\usepackage{appendix}
\usepackage{comment}
\usepackage{braket}
\usepackage{float}

\usepackage[colorlinks=true, allcolors=blue]{hyperref}
\usepackage[capitalise]{cleveref}

\usepackage{tikz}
\usetikzlibrary{calc}

\definecolor{pyred}{rgb}{0.839, 0.153, 0.157}
\definecolor{pyblue}{rgb}{0.122, 0.467, 0.705}

\newcommand{\dg}{^\dagger}
\newcommand{\hc}{\text{H.c.}}
\newcommand{\half}{\frac{1}{2}}

\newcommand{\atom}{a}
\newcommand{\res}{r}
\newcommand{\inter}{\text{int}}
\newcommand{\disp}{\text{disp}}
\newcommand{\tmon}{T}
\newcommand{\maj}{M}
\newcommand{\mtmon}{\text{MT}}
\newcommand{\mbx}{\text{MB}}
\newcommand{\tunn}{{t}}

\newcommand{\tl}{{t}}
\newcommand{\mt}{{mt}}
\newcommand{\mb}{{mb}}

\makeatletter
\AtBeginDocument{\let\LS@rot\@undefined}
\makeatother

\begin{document}

\title{Dispersive readout of Majorana qubits}

\author{Thomas~B.~Smith}
\affiliation{ARC Centre of Excellence for Engineered Quantum Systems, School of Physics, The University of Sydney, Sydney, NSW 2006, Australia.}

\author{Maja~C.~Cassidy}
\affiliation{Microsoft Quantum Sydney, The University of Sydney, Sydney, NSW 2006, Australia.}

\author{David~J.~Reilly}
\affiliation{ARC Centre of Excellence for Engineered Quantum Systems, School of Physics, The University of Sydney, Sydney, NSW 2006, Australia.}
\affiliation{Microsoft Quantum Sydney, The University of Sydney, Sydney, NSW 2006, Australia.}

\author{Stephen~D.~Bartlett}
\affiliation{ARC Centre of Excellence for Engineered Quantum Systems, School of Physics, The University of Sydney, Sydney, NSW 2006, Australia.}

\author{Arne~L.~Grimsmo}
\affiliation{ARC Centre of Excellence for Engineered Quantum Systems, School of Physics, The University of Sydney, Sydney, NSW 2006, Australia.}

\date{\today}

\begin{abstract} 
We analyze a readout scheme for Majorana qubits based on dispersive coupling to a resonator. We consider two variants of Majorana qubits: the Majorana transmon and the Majorana box qubit. In both cases, the qubit-resonator interaction can produce sizeable dispersive shifts in the MHz range for reasonable system parameters, allowing for submicrosecond readout with high fidelity. For Majorana transmons, the light-matter interaction used for readout manifestly conserves Majorana parity, which leads to a notion of quantum nondemolition (QND) readout that is stronger than for conventional charge qubits. In contrast, Majorana box qubits only recover an approximately QND readout mechanism in the dispersive limit where the resonator detuning is large. We also compare dispersive readout to longitudinal readout for the Majorana box qubit. We show that the latter gives faster and higher fidelity readout for reasonable parameters, while having the additional advantage of being manifestly QND, and so may prove to be a better readout mechanism for these systems.
\end{abstract}

\maketitle

\section{Introduction}\label{section:introduction}

\begin{figure*}
    \includegraphics{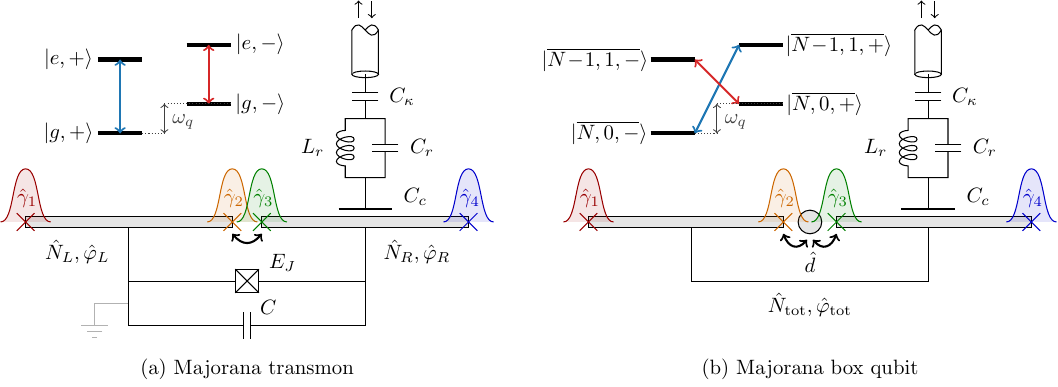}
	\caption{
	    Illustration of readout for two Majorana qubit variants. 
	    Topological superconductors (gray rectangles) host MZMs $\hat\gamma_i$, and qubit readout corresponds to a measurement of $i \hat\gamma_2 \hat\gamma_3$.
	    Majorana wavefunctions can be made to overlap either by (a) direct tunneling or (b) tunneling to a proximal quantum dot (gray circle). 
	    The resonator (illustrated by an LC oscillator) is capacitively coupled to the island charge. 
	    Alternative resonator-island coupling geometries are possible, including coupling to the quantum dot in (b). 
	    (a) Majorana transmon qubit: the topological superconductors form two distinct superconducting islands, shunted by a Josephson junction with energy $E_J$ (one of the islands may be grounded). 
	    The level diagram illustrates low-energy eigenstates labeled $\ket{g,\pm}$ and $\ket{e,\pm}$.
	    The blue and red arrows indicate allowed transitions induced by the qubit-resonator interaction. 
	    These allowed transitions conserve Majorana parity. 
	    (b) Majorana box qubit: the topological superconductors are shunted by a trivial superconductor, and the device forms a single superconducting island. 
	    The level diagram labels dressed eigenstates of the coupled superconducting island-dot system.
	    In this case, the resonator induces transitions between dressed states of different dot occupancy and Majorana parity.
	    }
	\label{figure:majorana-qubits}
\end{figure*}

Topological phases of matter offer a promising platform for quantum information processing, as qubits encoded into the degenerate ground states of these exotic phases should be extremely robust to errors~\cite{kitaev2003fault,bravyi2006universal,nayak2008non}. 
Engineered symmetry-protected topological phases have been proposed and recently investigated in one-dimensional hybrid semiconductor-superconductor nanowires~\cite{kitaev2001unpaired, fu2008superconducting, lutchyn2010majorana, oreg2010helical, sau2010generic, alicea2012new, mourik2012signatures, beenakker2013search, nadj2014observation, albrecht2016exponential, deng2016majorana, nichele2017scaling, lutchyn2018majorana,leijnse2012introduction, stanescu2013majorana}. 
Under specific conditions, these nanowires may enter a topological superconducting phase, distinguished by the emergence of Majorana zero-energy modes (MZMs). 
Fabricated devices with multiple nanowires could allow quantum computation with these MZMs~\cite{alicea2011non, sau2011controlling, van2012coulomb, hyart2013, sarma2015majorana, aasen2016milestones,plugge2017majorana, karzig2017scalable}.

Majorana-based quantum computing requires a measurement scheme for MZMs~\cite{ohm2015microwave, gharavi2016readout,li2018four, grimsmo2019majorana, Szechenyi2020,steiner2020readout, munk2020paritytocharge, maman2020charge, khindanov2020visibility}, which ideally should be fast and have high fidelity.
In measurement-only approaches to topological quantum computation, such measurements take on an especially important role, replacing braiding for the implementation of topologically protected quantum logic gates~\cite{bonderson2008measurement, bonderson2009measurement, plugge2017majorana, karzig2017scalable}.  Here, measurements must in addition be quantum nondemolition (QND). 
The QND property ensures that the measured observable is a conserved quantity, and constrains the postmeasurement state to be an eigenstate of the observable, such that repeated measurements give the same outcome. 
This is a critical requirement for measurement-only topological quantum computation with MZMs, where the measurements determine the dynamics of the system.
Beyond the implementation of topologically protected logic gates, measurements also play a key role in other aspects of Majorana-based quantum computing such as magic state distillation~\cite{karzig2019robust}.

Looking broadly at measurement schemes for solid-state quantum computing, a standard approach has been to couple a qubit-state-dependent charge dipole to the electric field of a resonator, which is used as a measurement probe~\cite{blais2020circuit}. 
A QND readout scheme for Majorana qubits based on parametric modulation of a qubit-resonator coupling was recently introduced in Ref.~\cite{grimsmo2019majorana}. 
However, the workhorse of measurement schemes for solid-state qubits is \emph{dispersive readout}, which has been very successful for superconducting~\cite{wallraff2004circuit, walter2017rapid}, semiconducting~\cite{colless2013dispersive, west2019gate, de2019rapid}, and hybrid semiconductor-superconductor qubits~\cite{larsen2015semiconductor, de2015realization}. 
In these schemes the resonator is tuned far off-resonance from the qubit frequency, and acquires a qubit-state-dependent frequency shift. 
It is natural to ask whether such a dispersive readout scheme can offer similar advantages for Majorana qubits, and to what extent it can satisfy the stringent QND requirements that are demanded by measurement-only topological quantum computation.

In this paper, we investigate a dispersive readout scheme for two prototypical Majorana qubits: the Majorana transmon~\cite{ginossar2014microwave, yavilberg2015fermion, hell2016time, aasen2016milestones} and the Majorana box qubit~\cite{plugge2017majorana, karzig2017scalable}. 
These two designs are distinguished by whether the two topological wire segments that host the four MZMs form two distinct superconducting charge islands or a single island with a uniform superconducting phase, respectively. 
We calculate the qubit-state-dependent dispersive shift that arises when these Majorana qubits are capacitively coupled to the electric field of a readout resonator. 
The size of these dispersive shifts directly determine the rate at which one can perform qubit readout by driving the resonator and observing the phase shift of the reflected field~\cite{blais2004cavity}. 
It also therefore determines the clock frequency in measurement-only approaches to quantum computation with MZMs.

Majorana qubits differ from conventional superconducting charge qubits, such as the Cooper pair box and the transmon~\cite{koch2007charge}, as a dispersive shift for a Majorana qubit can arise despite the fact that the interaction with the resonator does not induce (virtual) transitions between the two logical qubit states. 
Instead, the shifts result from (virtual) transitions to excited states outside the qubit subspace.
This is especially beneficial for the Majorana transmon, where dispersive shifts arise through a qubit-resonator interaction that conserves the Majorana charge parity.
For the Majorana box qubit, the Majorana parity is only approximately conserved in a limit of large frequency detuning from the resonator.

For both Majorana transmons and Majorana box qubits, we find that dispersive shifts can be comparable to those of conventional transmon qubits~\cite{walter2017rapid} and nanowire quantum dots~\cite{de2019rapid}. 
Specifically, for reasonable system parameters, we predict dispersive shifts in the megahertz range.
Measuring the qubit necessarily requires lifting the MZM degeneracy, and the corresponding qubit frequency is in the range $1$--$2$~GHz. 
Our results suggest that submicrosecond high-fidelity QND readout is feasible for Majorana qubits.

The remainder of the paper is organized as follows. 
We give a high-level introduction to dispersive coupling between a Majorana qubit and a resonator in~\cref{section:light-matter}. 
In~\cref{section:majorana-transmon,section:majorana-box} we describe in detail the dispersive coupling for a Majorana transmon and a Majorana box qubit, respectively. 
For both qubit variants, we calculate the dispersive frequency shift of a readout resonator from second-order Schrieffer-Wolff perturbation theory for a range of system parameters using numerical diagonalization. 
We also provide simple, approximate analytical expressions. 
We estimate the resulting measurement timescales and fidelities (in the absence of any unwanted qubit decoherence or noise in the system parameters; see e.g.,~\cite{knapp2018dephasing, karzig2020quasiparticle, khindanov2020visibility, maman2020charge}) in~\cref{section:readout-estimates}.
We compare the results for dispersive readout to the longitudinal readout scheme introduced in Ref.~\cite{grimsmo2019majorana}.
Finally, in~\cref{section:conclusions} we discuss the implications of our results for dispersive readout of Majorana qubits.

\section{Light-matter interaction for Majorana qubits}\label{section:light-matter}

We begin by presenting a high-level overview of the interaction between a Majorana qubit and an electromagnetic resonator, focusing on the dispersive coupling regime.
Such a coupling provides the underlying physical mechanism that can be used for dispersive qubit readout~\cite{blais2020circuit}. 

The resonator-based readout schemes considered in this paper involve capacitive coupling of the charge degree of freedom of the measured system (the qubit) to the electric field of a nearby resonator. 
That is, we have an interaction Hamiltonian of the form 
\begin{equation}\label{equation:light-matter-interaction}
    \hat H_\inter = \hat{Q} \hat{V}_\res = i \hbar \lambda \hat N (\hat a\dg - \hat a),
\end{equation}
where $\hat Q= e \hat N$ is a charge operator for the qubit system, $\hat V_r \sim i(\hat a\dg -  \hat a)$ is the voltage bias on the qubit due to the resonator, and $\lambda$ quantifies the interaction strength. 
For simplicity, we model the resonator by a single harmonic oscillator mode with annihilation (creation) operator $\hat a$~($\hat a\dg$).

The physics resulting from the coupling described by~\cref{equation:light-matter-interaction} depends on the internal level structure of the qubit system. 
Given that the charge number operator $\hat N$ can have off-diagonal matrix elements in the qubit eigenbasis, the absorption or emission of a resonator photon can induce a transition between eigenstates in the qubit system. 
The internal level structure of the qubit can lead to selection rules where only certain transitions are allowed. 
We show below that different Majorana qubit designs give rise to different selection rules, and discuss the consequences of this for QND readout.

We consider two distinct types of Majorana qubits, illustrated in~\cref{figure:majorana-qubits}. 
Each topological superconductor hosts a pair of Majorana edge modes, and we associate to each edge mode a Hermitian Majorana fermion operator $\hat \gamma_i$, with anticommutation relations $\{\hat \gamma_i, \hat \gamma_j\} = 2\delta_{ij}$.
Because of charge conservation, a minimum of two topological superconductors are required to encode a Majorana qubit, for a total of four MZMs.
We identify two broad classes of Majorana qubits, depending on whether these topological superconductors form one or two distinct superconducting charge islands. 
The Majorana transmon qubit is representative of a configuration where the two topological superconductors form two distinct islands, and the relevant charge degree of freedom for readout is the difference in charge between these two islands. 
Variations of this configuration can include grounding one of the two superconducting islands, and/or introducing a Josephson tunnel coupling between the islands and ground~\cite{aasen2016milestones}. 
However, it should be noted that connecting one of the islands to ground in this manner could increase the rate of quasiparticle poisoning events~\cite{karzig2020quasiparticle}.
For the Majorana box qubit (also referred to as the Majorana loop qubit), the two topological superconductors are shunted by a trivial superconductor to form a single superconducting island. 
In this case, a charge dipole can be formed by tunnel coupling to a proximal quantum dot, providing a mechanism for readout.

The physics of these devices is described in more detail in the following sections, and we here only give a high-level discussion of their internal level structure and selection rules. 
For the Majorana transmon in~\cref{figure:majorana-qubits}~(a), energy levels can be labeled $\ket{g,\pm}, \ket{e,\pm}, \dots$, where $g,e,\dots$ denote a transmonlike ladder of eigenstates, and $\pm$ denotes the eigenvalue of $i \hat\gamma_2 \hat\gamma_3 = \pm 1$, the Majorana parity we wish to measure. 
As indicated in the level diagram in~\cref{figure:majorana-qubits}~(a), only transitions that conserve the Majorana parity are allowed. 
This means that the Majorana parity is conserved during readout and that the interaction is manifestly QND with respect to this quantity. 
This stronger-than-usual form of QND measurement stems from the fractional and nonlocal nature of the MZMs~\cite{plugge2017majorana}, and was dubbed \textit{topological QND} (TQND) measurement in Ref.~\cite{grimsmo2019majorana}. 
This TQND property plays a key role in maintaining the topological protection of measurement-based approaches to implementing quantum logic gates.
 
For the Majorana box qubit, MZMs are tunnel coupled to a proximal quantum dot, as illustrated in~\cref{figure:majorana-qubits}~(b).
The dot might be formed naturally between two topological superconductors due to the boundary conditions set by the superconducting-semiconducting interface~\cite{deng2018nonlocality}.
In our analysis we assume that this quantum dot has well-separated energy levels, and for simplicity only a single level that is energetically accessible.

Charge tunneling between the topological superconducting island and the dot provides a mechanism for readout.
Because the superconducting island charge is no longer conserved, the eigenstates of the qubit-dot system are dressed states where the Majorana edge modes are partially localized on the dot. 
These dressed states are illustrated in the level diagram in~\cref{figure:majorana-qubits}~(b).
In a readout protocol, the tunnel coupling should be turned on gradually, such that the system evolves adiabatically from the bare to the dressed eigenstates, and we label the dressed eigenstates by the states they are adiabatically connected to in the absence of tunneling. 
Qubit readout corresponds to distinguishing the dressed states adiabatically connected to the degenerate qubit ground space.

As indicated in the level diagram in~\cref{figure:majorana-qubits}~(b), the relevant transitions for coupling to the resonator involve a single charge transfer from the island to the dot, or vice versa. 
This charge transfer also flips the (dressed) Majorana parity. 
In this case, an approximately QND interaction can still be achieved in the dispersive regime, where the resonator is far detuned from any internal transition, such that the resonator-induced transitions indicated in~\cref{figure:majorana-qubits}~(b) are purely virtual. 
However, it is a notable difference between the Majorana transmon and the Majorana box qubit readout scheme that the former has the advantage of a manifestly QND interaction independent of whether the system is in the dispersive regime or not.

It is worth noting that the \emph{joint} parity of the Majorana box qubit and the dot is, in fact, preserved by the transitions indicated in~\cref{figure:majorana-qubits}~(b). One can therefore perform a QND measurement of this joint parity, and in principle infer the Majorana parity, given that the state of the dot can be determined with certainty before and after the measurement~\cite{steiner2020readout, munk2020paritytocharge}.

As an aside, the lack on an exact QND interaction for the dispersive readout of the Majorana box qubit can be contrasted to the longitudinal readout scheme proposed in Ref.~\cite{grimsmo2019majorana}, where modulation of a system parameter is used to activate a parity conserving qubit-resonator coupling. 
This coupling arises \emph{independently} of the frequency detuning of the resonator from any internal qubit transition. 
Therefore, this longitudinal readout scheme can be used in a regime where the detuning is large enough that any Majorana parity nonconserving terms are negligible, yet still achieving a large readout signal-to-noise ratio.
We return to a brief comparison with Ref.~\cite{grimsmo2019majorana} in~\cref{section:readout-estimates,appendix:majorana-box-longitudinal}.

For the purpose of qubit readout, real transitions between qubit-system eigenstates are undesirable. 
An effective interaction suitable for readout is recovered from~\cref{equation:light-matter-interaction} in the dispersive regime. 
This refers to a coupling regime where the resonator frequency is far off-resonance from any relevant transitions between qubit states that are allowed by the selection rules. 
The transitions to higher energy levels indicated in~\cref{figure:majorana-qubits}~(a,b) are then only virtual transitions.
In this situation,~\cref{equation:light-matter-interaction} can be treated perturbatively, leading to an effective interaction of the form (for both types of Majorana qubit)
\begin{equation}\label{equation:dispersive-interaction}
	\hat H_\disp = \hbar \omega_\res \hat a\dg \hat a + \frac{\hbar \omega_q}{2} \hat \sigma_z + \hbar \chi_q \hat \sigma_z \hat a\dg \hat a.
\end{equation}
Here $\omega_\res$ is the resonator frequency, $\hbar \omega_q$ is the energy splitting between the two eigenstates used to encode a qubit, and $\hat\sigma_z$ is the corresponding logical Pauli-$Z$ operator. 
In general, $\omega_{\res,q}$ include Lamb shifts due to the qubit-resonator coupling. 
Finally, $\chi_q$ is the qubit-state-dependent dispersive frequency shift of the resonator. 
Under this Hamiltonian, the qubit states can be distinguished by detecting a phase shift of the resonator under a coherent drive at the resonator frequency~\cite{blais2004cavity, wallraff2004circuit, koch2007charge, blais2020circuit}. 
The speed of such a measurement is set by the magnitude of the dispersive shift $\chi_q$. 
We give a derivation of~\cref{equation:dispersive-interaction} starting from~\cref{equation:light-matter-interaction}, for a generic multilevel system, in~\cref{appendix:dispersive-regime-derivation}.
Throughout this paper, we compute $\chi_q$ for three different qubit types labeled $q \in \{\tl,\mt,\mb\}$, for a conventional transmon, a Majorana transmon, and a Majorana box qubit, respectively.

It is important to emphasize that although~\cref{equation:dispersive-interaction} is QND with respect to the logical $\hat\sigma_z$ operator, this Hamiltonian is an approximation to the underlying light-matter interaction,~\cref{equation:light-matter-interaction}. 
The TQND property of the Majorana transmon refers to the fact that parity protection is manifest at the more fundamental level of~\cref{equation:light-matter-interaction}. 
As discussed briefly above, and in more detail in the following, dispersive readout for the Majorana box qubit is not TQND in the same strong sense as for the Majorana transmon. 
Both the Majorana transmon and the Majorana box qubit, however, share the feature that no transitions are allowed between the two lowest energy eigenstates used to form a qubit. 
Instead (virtual) transitions out of the qubit subspace are used to realize a readout mechanism.  
This is in contrast to conventional superconducting charge qubits, such as the transmon and the Cooper pair box, where the light-matter interaction causes transitions between the energy eigenstates that define the qubit~\cite{koch2007charge}. 
In this case, the readout mechanism introduces a source of error in the form of Purcell decay, wherein the qubit may relax via emission of a photon via the resonator~\cite{houck2008controlling}.
(We note that we have restricted our notion of measurement back action to the readout mechanism itself.
Additional unwanted effects that may be introduced such as quasiparticle poisoning~\cite{karzig2020quasiparticle} or heating are not treated here.)

In the following sections, we describe the dispersive readout schemes for the Majorana transmon and the Majorana box qubit in detail.

\section{Majorana transmon qubit}\label{section:majorana-transmon}

\subsection{Model for the qubit}\label{subsection:majorana-transmon-model}

A Majorana transmon qubit, shown in~\cref{figure:majorana-qubits}~(a), consists of two distinct charge islands that are shunted by a Josephson junction. 
Each island, labeled $\alpha \in \{L,R\}$, is in a topological superconducting phase and has electron number operator $\hat{N}_{L,R}$ and dimensionless superconducting phase operator $\hat\varphi_{L,R}$, satisfying $[\hat N_\alpha, e^{i \hat\varphi_\beta/2}] = \delta_{\alpha\beta} e^{i \hat\varphi_\beta/2}$, with $\alpha, \beta \in \{L,R\}$. 
The charging energy and conventional Cooper pair tunneling between the two islands is captured by a Hamiltonian
\begin{equation}\label{equation:transmon}
	\hat{H}_\tmon = E_C \big(\hat{N} - n_g \big)^2 - E_J \cos{\hat{\varphi}},
\end{equation}
where $\hat N \equiv (\hat N_L - \hat N_R)/2$ and $\hat\varphi \equiv \hat\varphi_L - \hat\varphi_R$, $E_C$ is the charging energy due to capacitive coupling of the two islands, $n_g$ represents an offset charge, and $E_J$ is the Josephson coupling due to Cooper pair tunneling across the Josephson junction. 
The transmon regime is characterized by $E_J \gg E_C$~\cite{koch2007charge}. 
Note that here we use a convention where $\hat N$ counts the number of electrons rather than the number of Cooper pairs, such that we have the following action on charge eigenstates:
\begin{subequations}\label{equation:charge-basis}
	\begin{align}
		\hat N \ket{N} &= N \ket{N},
		\\ e^{\pm i \hat\varphi} \ket{N} &= \ket{N \pm 2}.
	\end{align}
\end{subequations}
We have neglected the capacitances of each superconducting island to ground, and assume the long-island limit where MZMs located on the same island are well separated.

Variations of the Majorana transmon include grounding one of the two islands (such that we can set e.g., $\hat\varphi_L = 0$ and $\hat\varphi = \hat\varphi_R$) and/or introducing a Josephson coupling to a bulk superconductor in addition to the Josephson coupling between the two islands~\cite{aasen2016milestones}. 
These variations are qualitatively similar, and our results extend to these cases without any significant modification.

To read out this qubit, the MZMs corresponding to $\hat\gamma_2$ and $\hat\gamma_3$ are brought together and the combined parity $i \hat\gamma_2 \hat\gamma_3$ is measured. 
When these two MZMs are brought together (see~\cref{figure:majorana-qubits}), their interaction is governed by a tunneling Hamiltonian~\cite{kitaev2001unpaired, ginossar2014microwave, hell2016time, aasen2016milestones}
\begin{equation}\label{equation:majorana-transmon-tunneling}
	\hat{H}_\maj = - E_M {i}\hat\gamma_2 \hat\gamma_3 \cos{\left( \frac{\hat\varphi + \varphi_x}{2} \right)},
\end{equation}
where $E_M$ is proportional to the wavefunction overlap of the MZMs.
This expression accounts for the fact that the qubit loop might enclose an external flux $\Phi_x$, where in we have defined $\varphi_x = 2\pi\Phi_x/\Phi_0$ with $\Phi_0 = h/2e$ the magnetic flux quantum. 
In~\cref{appendix:majorana-interaction-models} we compare the direct tunneling model~\cref{equation:majorana-transmon-tunneling} with a model where the two islands are coupled to a common quantum dot, acting as a mediator.
The main outcome of this comparison is that, when the energy penalty to occupy the quantum dot becomes large, the two models are equivalent.

\begin{figure}
	\includegraphics[width=\linewidth]{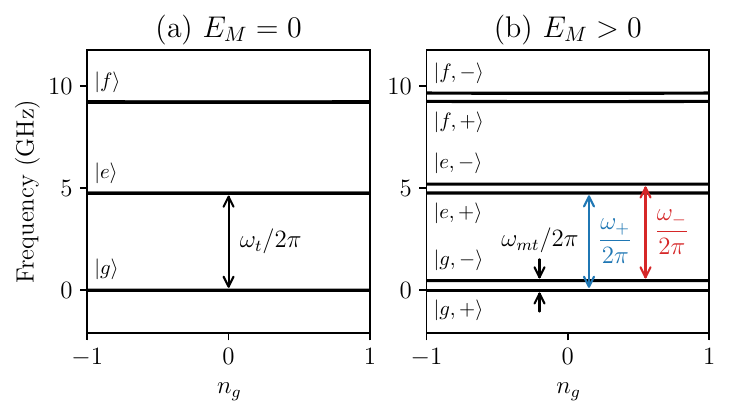}
	\caption{
		The low-level spectrum of a Majorana transmon qubit as a function of the offset charge parameter $n_g$, where $E_C/h = 250$ MHz, $E_J/E_C = 50$ ($\omega_\tl/2\pi \simeq 5$ GHz), and (a) $E_M = 0$ and (b) $E_M/\hbar\omega_t = 0.05$ ($\omega_\mt/\omega_t \simeq 0.1$). 
		When $E_M=0$, each level in the transmon spectrum is two-fold degenerate.
		This degeneracy is lifted when $E_M > 0$. 
		The resulting level structure can be thought of as two parallel transmon ladders, for $i \hat\gamma_2 \hat\gamma_3 = \pm 1$.
	    }
	\label{figure:majorana-transmon-spectra}
\end{figure}

The full Majorana transmon qubit Hamiltonian is
\begin{equation}\label{equation:majorana-transmon}
	\hat H_\mtmon = \hat H_\tmon + \hat H_\maj.
\end{equation}
Since~$i \hat\gamma_2 \hat\gamma_3$~commutes with~$\hat H_\mtmon$, the eigenstates can conveniently be labeled by two quantum numbers~$\ket{j, a}$ where~$j = g,e,f, \dots$ denote a transmonlike ladder of eigenstates and $a = \pm$ denotes the eigenvalue of $i \hat\gamma_2 \hat\gamma_3 = \pm 1$. 
The level structure is shown in~\cref{figure:majorana-transmon-spectra} for (a) $E_M = 0$ and (b) $E_M > 0$.

A simplified Hamiltonian can be found by following the standard approach of treating the transmon degree of freedom as a Kerr nonlinear oscillator~\cite{koch2007charge,blais2020circuit}.
To keep the discussion simple, we set the offset charge and external flux to zero, $n_g = 0$, $\varphi_x=0$, for the remainder of this section.
We can introduce ladder operators via 
\begin{subequations}\label{equation:ladder-operators}
	\begin{align}
		\hat N &= {i} \left( \frac{E_J}{2E_C} \right)^{1/4} \big(\hat b\dg - \hat b\big),
	    \\ \hat\varphi &= \left( \frac{2E_C}{E_J} \right)^{1/4} \big(\hat b\dg + \hat b\big).
	\end{align}
\end{subequations}
where $[\hat b, \hat b\dg] = 1$.
Taylor expanding the $\cos{\hat\varphi}$ term to fourth order in $\hat\varphi$, substituting the expressions above, and dropping fast rotating terms, we obtain the standard result
\begin{equation}\label{equation:transmon-effective}
	\hat H_\tmon \simeq \hbar\omega_\tl \hat b\dg \hat b - \frac{E_C}{2} \hat b\dg \hat b\dg \hat b \hat b,
\end{equation}
where $\hbar\omega_\tl = \sqrt{8E_JE_C} - E_C$ is the transmon energy and $E_C$ is the anharmonicity.

Repeating the steps for the Majorana term $\hat H_\maj$ yields
\begin{equation}\label{equation:majorana-transmon-tunneling-effective}
	\hat H_\maj \simeq -{i} E_M \hat\gamma_2 \hat\gamma_3 \left(1 - \xi_0 - \xi_1 \hat b\dg \hat b + \xi_2 \hat b\dg \hat b\dg \hat b \hat b \right),
\end{equation}  
with coefficients
\begin{subequations}\label{equation:majorana-transmon-tunneling-effective-coefficients}
	\begin{align}
		\xi_0 &= \frac{4\sqrt{8E_JE_C} - E_C}{64E_J},
		\\ \xi_1 &= \frac{4\sqrt{8E_JE_C} - 2E_C}{32E_J},
		\\ \xi_2 &= \frac{E_C}{32 E_J}.
	\end{align}
\end{subequations}
Under the above approximations, we see that the energy splitting of the two lowest energy levels with unequal Majorana parity, $\ket{g,\pm}$, which we label $\hbar\omega_\mt$, is
\begin{equation}\label{equation:majorana-transmon-qubit-splitting}
    \hbar\omega_\mt \simeq 2E_M(1-\xi_0).
\end{equation}
On the other hand, the ``transmon transitions'' $\ket{g,+} \leftrightarrow \ket{e,+}$ and $\ket{g,-} \leftrightarrow \ket{e,-}$ (indicated in~\cref{figure:majorana-transmon-spectra}) have energy splittings
\begin{equation}\label{equation:majorana-transmon-transition-splittings}
	\hbar\omega_\pm \simeq \hbar\omega_\tl \pm E_M\xi_1,
\end{equation}
respectively. 
As we will show, despite that there is no charge matrix element between the two logical qubit states $\ket{g,\pm}$, the fact that the two transition frequencies $\omega_\pm$ are nondegenerate for $E_M > 0$ nevertheless leads to a $i \hat\gamma_2 \hat\gamma_3$-dependent dispersive shift of the resonator.

\subsection{Dispersive interaction with a resonator}\label{subsection:majorana-transmon-dispersive-regime}

The Majorana transmon qubit can be read out via a resonator that is capacitively coupled to the island charge, as schematically illustrated in~\cref{figure:majorana-qubits}. 
This interaction has the form
\begin{equation}\label{equation:majorana-transmon-resonator-interaction}
	\hat H_\inter = i\hbar\lambda \hat N (\hat a\dg - \hat a),
\end{equation}
while the resonator Hamiltonian is given by $\hat H_\res = \hbar\omega_\res \hat a\dg \hat a$. 
Here $\hbar \lambda \simeq 2 (C_c/C_\res) E_C \sqrt{R_K/4\pi Z_\res}$ quantifies the capacitive coupling strength, with $C_c$ the coupling capacitance, $C_\res$ the resonator capacitance, $Z_\res = \sqrt{L_\res/C_\res}$ the resonator characteristic impedance, and $R_K = h/e^2$ the resistance quantum. 

We numerically diagonalize the full Hamiltonian $\hat H_\mtmon$,~\cref{equation:majorana-transmon}, to calculate the dispersive shift $\chi_\mt$ defined in~\cref{equation:qubit-dipserive-shift-list}. 
To assist with interpreting our results, we first calculate dispersive shifts for a conventional transmon qubit, which corresponds to the limit $E_M = 0$. 

\begin{figure}
	\includegraphics[width=\linewidth]{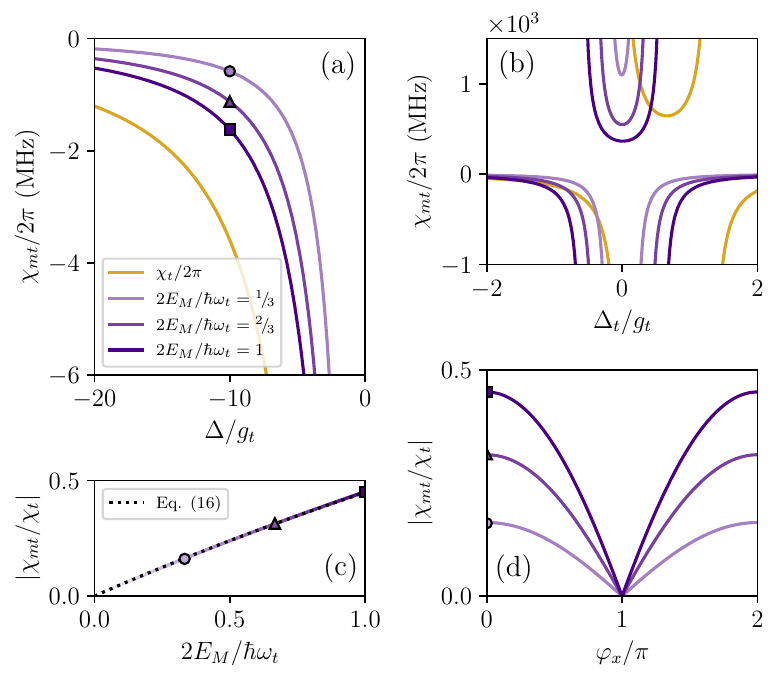}
	\caption{
		Dispersive frequency shift of a readout resonator coupled to a Majorana transmon qubit $\chi_\mt$ (purple) in comparison to those of a conventional transmon qubit $\chi_\tl$ (gold), where $E_C/h = 250$ MHz, $E_J/E_C = 50$ ($\omega_\tl/2\pi \simeq 5$ GHz), $n_g = 0$, and $\lambda/2\pi = 100$ MHz ($g_\tl/2\pi \simeq 200$ MHz). 
		In (a), (b), (c) $\varphi_x=0$ and in (b), (c) $\Delta/g_\tl=-10$. 
		Circle, triangle and square markers indicate equivalent points between plots. 
		The top plots show $\chi_\mt$ and $\chi_\tl$ as a function of (a) $\Delta/g_\tl$ and (b) $\Delta_t/g_\tl$ for three values of $E_M$ where $\varphi_x=0$. 
		In (a) $\Delta$ is the detuning from the relevant transition frequency ($\omega_t$ for the transmon and $\omega_+$ for the Majorana transmon). 
		In (b) $\Delta_\tl = \omega_\tl - \omega_\res$, plotted such that the saddle regimes are visible. 
		The magnitude of $\chi_\mt$ in comparison to $\chi_\tl$ is plotted as a function of (c) $2E_M/\hbar\omega_\tl$ ($\simeq \omega_\mt/\omega_\tl$) and (d) $\varphi_x/\pi$.
	    }
    \label{figure:majorana-transmon-dispersive-shift}
\end{figure}

\textit{Conventional transmon.}---In this case, the qubit is encoded in the two eigenstates $\ket 0 \equiv \ket{g,a}$ and $\ket 1 \equiv \ket{e,a}$ where the choice of $a = \pm$ is arbitrary.
The spectrum is shown in~\cref{figure:majorana-transmon-spectra}~(a). 
There are two primary transitions that contribute to the conventional transmon dispersive shift $\chi_\tl$, defined in~\cref{equation:qubit-dipserive-shift-list}.
Namely, the qubit transition $\ket{g,a} \leftrightarrow \ket{e,a}$, with frequency $\omega_\tl$, and the transition $\ket{e,a} \leftrightarrow \ket{f,a}$ with frequency approximately given by $\omega_\tl - E_C/\hbar$. 
We numerically calculate $\chi_\tl$ as a function of the detuning parameter $\Delta_t \equiv \omega_\tl - \omega_\res$ in~\cref{figure:majorana-transmon-dispersive-shift}~(b). 
The singularities at $\Delta_t = 0$ and $\Delta_t = E_C/\hbar$ correspond to values of $\omega_\res$ where the resonator is resonant with the $\ket{g,a}\leftrightarrow \ket{e,a}$ and $\ket{e,a} \leftrightarrow \ket{f,a}$ transitions, respectively. 
The regime between these two singularities, where the dispersive shift changes sign, is known as the straddling regime~\cite{koch2007charge}. 
The transmon dispersive shift $\chi_\tl$ provides a point of comparison that we use to evaluate the dispersive shifts of Majorana transmons.

\textit{Majorana transmon.}---We next consider the Majorana transmon qubit with $E_M > 0$. 
For the Majorana transmon qubit, the logical qubit states are chosen to be $\ket 0 \equiv \ket{g,+}$ and $\ket 1 \equiv \ket{g,-}$ instead. 
The qubit-resonator interaction does not induce transitions between levels with different parity $a$.
Instead, the contributions to the qubit-state-dependent dispersive shift $\chi_\mt$, in analogy to $\chi_\tl$, come from transitions $\ket{g,+} \leftrightarrow \ket{e,+}$ with frequency $\omega_+$, and $\ket{g,-} \leftrightarrow \ket{e,-}$ with frequency $\omega_-$, as indicated in~\cref{figure:majorana-transmon-spectra}. 
The frequencies $\omega_\pm$ are close to $\omega_\tl$, approximately given by~\cref{equation:majorana-transmon-transition-splittings}. 
As $2E_M \simeq \hbar\omega_\mt$ increases and becomes comparable to $\hbar\omega_\tl$, $\chi_\mt$ approaches a comparable magnitude to $\chi_\tl$, the dispersive shift for a conventional transmon, as shown in~\cref{figure:majorana-transmon-dispersive-shift}~(a,c).
We note that the strength of the dispersive shift $\chi_\mt$ also depends on the offset flux $\varphi_x$, as shown in~\cref{figure:majorana-transmon-dispersive-shift}~(d).
Care must be taken to ensure that $\varphi_x \neq \pi$, where $\chi_\mt$ vanishes and changes sign.

We can also find an approximate analytical expression for the dispersive shift. 
To this end, we substitute \cref{equation:ladder-operators} into $\hat H_\inter$ and make a rotating-wave approximation to find that
\begin{equation}\label{equation:majorana-transmon-resonator-interaction-effective}
	\hat H_\inter \simeq \hbar g_\tl (\hat b \hat a\dg + \hat a \hat b\dg),
\end{equation}
where
\begin{equation}\label{equation:transmon-resonator-coupling}
    g_\tl = \lambda \left( \frac{E_J}{2E_C} \right)^{1/4}.
\end{equation}
This form clearly shows how energy exchange with the resonator leads to transitions between transmon levels within the same parity sector ($\ket{g,\pm} \leftrightarrow \ket{e,\pm}$, $\ket{e,\pm} \leftrightarrow \ket{f,\pm}$, etc.).
From~\cref{equation:majorana-transmon-resonator-interaction-effective} we find the following simple approximate expression for $\chi_\mt$ (see~\cref{appendix:majorana-transmon-schrieffer-wolff}):
\begin{equation}\label{equation:majorana-transmon-dispersive-shift-effective}
	\chi_\mt \simeq \frac{1}{2}\left(\frac{g_\tl^2}{\omega_{+}-\omega_\res} - \frac{g_\tl^2}{\omega_{-}-\omega_\res}\right).
\end{equation}
This is compared to the result based on exact diagonalization of the qubit Hamiltonian in \cref{figure:majorana-transmon-dispersive-shift}~(c). 
We emphasize that~\cref{equation:majorana-transmon-dispersive-shift-effective} is a somewhat crude approximation similar in accuracy to the standard approximation used for conventional transmon qubits [see Eq. (3.12) of Ref.~\cite{koch2007charge}].

\section{Majorana box qubit}\label{section:majorana-box}

\subsection{Model for the qubit}\label{subsection:majorana-box-model}

The Majorana box qubit, shown in~\cref{figure:majorana-qubits}~(b), is an alternative design for a qubit based on MZMs that has been studied in the context of measurement-only topological quantum computing~\cite{plugge2017majorana, karzig2017scalable}. 
In this qubit design, the topological superconductors are shunted by a (trivial) superconducting bridge instead of a Josephson junction. 
This model can be thought of as a limiting case to the Majorana transmon, where $E_J/E_C \rightarrow \infty$. 
In this limit, we have $\hat\varphi_R \to \hat\varphi_L$, such that the previous charge and phase operators, $\hat N = (\hat N_L - \hat N_R)/2$ and $\hat\varphi = \hat\varphi_L - \hat\varphi_R$, are zero. 
Instead, the relevant degree of freedom is the total charge $\hat N_\text{tot} = \hat N_L + \hat N_R$, and the corresponding conjugate phase $\hat\varphi_\text{tot} \equiv (\hat\varphi_L + \hat\varphi_R)/2$. 
The device acts as a single island with charging energy
\begin{equation}\label{equation:majorana-box-charge}
	\hat{H}_\mathrm{tot} = E_\mathrm{tot} ( \hat{N}_\mathrm{tot} - n_g )^2,
\end{equation}
where $E_\mathrm{tot}$ quantifies the charging energy due to capacitive coupling of the island to ground (and to the resonator), which we neglected for the Majorana transmon. 
The charge and phase operators $\hat N_\text{tot}$, $\hat \varphi_\text{tot}$ act on charge eigenstates analogously to~\cref{equation:charge-basis}, where $\hat N_\text{tot}$ now counts the total charge on the superconducting island consisting of the two topological superconductors.

The Majorana box qubit also comes in several qualitatively similar variations. 
When the two topological superconductors are aligned horizontally in series as in~\cref{figure:majorana-qubits}~(b) (formed from a single nanowire), the qubit is also refereed to as a Majorana loop qubit. 
Alternatively, the two topological superconductors can be arranged in parallel with the superconducting shunt perpendicular to the nanowires~\cite{plugge2017majorana}.
One can also consider additional MZMs per island, used as ancilla modes for measurement-only topological quantum computing. 
In this case, a Majorana box qubit with four MZMs is called a tetron, with six MZMs a hexon, and so on. 
Our results can be generalized to these variations.

The coupling of the two topological superconductors due to a nonzero overlap of the Majorana modes corresponding to $\hat\gamma_2$ and $\hat\gamma_3$ can be modeled using~\cref{equation:majorana-transmon-tunneling} with $\hat \varphi\to 0$. 
However, to properly account for the movement of charge that leads to a coupling to the resonator, we take one step back and explicitly include coupling to bound states in the semiconducting region between the two topological superconductors. 
In the limit of where the energy penalty to occupy these bound states is large compared to the tunnel coupling, the Hamiltonian $\hat H_\maj$ can be recovered as an effective description.
Including such bound states as intermediate degrees of freedom is however necessary to correctly capture the coupling to the resonator that results from charge tunneling to the semiconductor. 
In the proposals in Refs.~\cite{plugge2017majorana, karzig2017scalable}, this description is moreover very natural because a gate-defined quantum dot is explicitly introduced to mediate a tunable interaction between the nanowires.

We model the quantum dot between the two topological superconductors by a single fermionic operator $\hat d$, satisfying $\{\hat d,\hat d\dg\} = 1$. This degree of freedom is illustrated in~\cref{figure:majorana-qubits}~(b). 
The dot is described by a Hamiltonian $\hat H_d = \varepsilon \hat d\dg \hat d$, with $\varepsilon$ the dot occupation energy. Tunneling between the island and the dot is modeled by a Hamiltonian~\cite{fu2010electron}
\begin{equation}\label{equation:majorana-box-tunneling}
	\hat{H}_\tunn = \frac{1}{2}\left[ e^{\frac{i\hat{\varphi}_\mathrm{tot}}{2}} \left( it_Le^\frac{i\varphi_x}{2} \hat{\gamma}_2 - t_R \hat{\gamma}_ 3\right)\hat{d} + \mathrm{H.c.} \right],
\end{equation}
where $t_{L,R} \ge 0$ are the tunneling amplitudes between the two respective  topological superconductors, and we have included the possibility of an external flux, $\varphi_x$, threading the qubit loop. 
The full Majorana box qubit Hamiltonian is thus
\begin{equation}\label{equation:majorana-box}
	\hat H_\mbx = \hat H_\mathrm{tot} + \hat H_d + \hat H_\tunn.
\end{equation}

\begin{figure}
	\includegraphics[width=\linewidth]{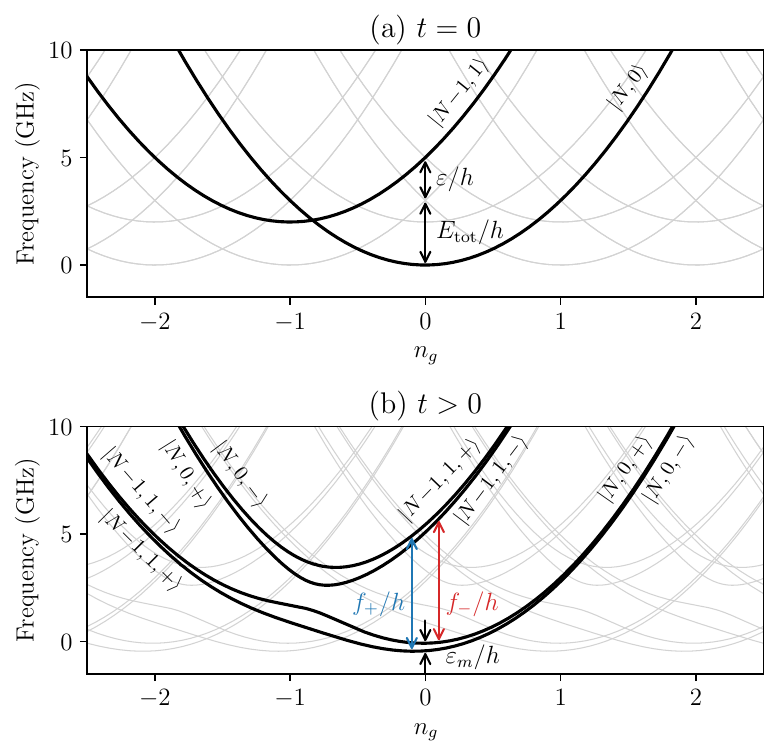}
	\caption{
    	The low-level spectrum of a Majorana box qubit as a function of the offset charge parameter $n_g$, where $\delta/h = (E_\mathrm{tot}+\varepsilon)/h = 5$ GHz, $\varphi_x = \pi/2$, $t_{L,R} = t$, and (a) $t = 0$ and (b) $t/\delta \simeq 0.2$ ($\varepsilon_m/h \simeq 0.5$ GHz). 
    	When MZMs interact via the quantum dot $(t > 0)$, the charge state $|N, n_d=0, \pm \rangle$ hybridizes with $|N-1, n_d=1, \mp \rangle$.
    	}
	\label{figure:majorana-box-spectra}
\end{figure}

We show the spectrum in the uncoupled case $t_{L,R} = 0$ in \cref{figure:majorana-box-spectra}~(a).
As with the Majorana transmon qubit, each state shown in \cref{figure:majorana-box-spectra}~(a) is two-fold degenerate, a degeneracy that splits when we include nonzero tunneling $|t_{L,R}| > 0$, as shown in \cref{figure:majorana-box-spectra}~(b).

The dot-island Hamiltonian $\hat H_\mbx$ conserves the total charge $\hat N_\text{tot} + \hat d\dg \hat d$, such that the Hamiltonian can be diagonalized block by block, following Ref.~\cite{grimsmo2019majorana}. 
After a unitary transformation $\hat H_\mbx' = \hat U\dg \hat H_\mbx \hat U$ we have
\begin{equation}\label{equation:majorana-box-diagonal}
    \begin{aligned}
        \hat H'_\mbx
        ={}&\varepsilon_{c}(\hat N+1) \hat d\dg \hat d
        + \frac{\varepsilon_m(\hat N)}{2}i\hat \gamma_2\hat\gamma_3 \\
        &+ \half \left[\varepsilon_m(\hat N+1)-\varepsilon_m(\hat N)\right]\hat d\dg \hat d
        \,\hat i\hat\gamma_2\hat\gamma_3 \\
        &+ E(\hat N) + \left[E(\hat N+1)- E(\hat N)\right]\hat d\dg \hat d,
    \end{aligned}
\end{equation}
The functions $\varepsilon_c(n)$, $\varepsilon_m(n)$ and $E(n)$ are given in~\cref{equation:majorana-box-diagonal-coefficients}.

As is clear from~\cref{equation:majorana-box-diagonal}, the eigenstates of $\hat H'_\mbx$ can be labeled by three quantum numbers $\ket{N,n_d,a}$: the island charge $N \in \mathbb{Z}$, the dot occupancy $n_d=0,1$, and the Majorana parity $a = \pm$.
The dressed eigenstates of $\hat H_\mbx$, \cref{equation:majorana-box}, are thus related to the bare charge states of the uncoupled system (i.e., when $t_{L,R} = 0$) through
\begin{equation}
    \overline{\ket{N, n_d, a}} = \hat U \ket{N, n_d, a},
\end{equation}
where the unitary transformation is defined in~\cref{appendix:majorana-box-diagonalization}. 
The labels on the left-hand side here designate hybridized degrees of freedom, in particular, when $t_{L,R}>0$, the Majorana fermion hybridizes with the dot, and $a = \pm$ refers to the corresponding ``dressed'' Majorana parity.

To keep the discussion simple, from now on we focus on the sector with zero total dot-island charge, $\hat N + \hat d\dg \hat d = 0$, and set $t_{L,R} = t$, $n_g = 0$ for the remainder of this section. Within the zero total charge sector, \cref{equation:majorana-box-diagonal} takes the form
\begin{equation}\label{equation:majorana-box-diagonal-0}
	\hat H_{\mbx,0} = \varepsilon_c \hat{c}^\dagger \hat{c}	+ \frac{\varepsilon_m}{2}i\hat{\gamma}_2\hat{\gamma}_3,
\end{equation}
where we have dropped a constant term, and $\hat c$ satisfies $\{\hat c, \hat c\dg\} = 1$ and 
describes the movement of an electron from the dot to the island within the zero total charge sector.
The coefficients are given by
\begin{subequations}\label{equation:majorana-box-diagonal-coefficients-0}
	\begin{align}
		\varepsilon_c &= \frac{1}{2}\left(f_+ + f_- \right),\\
		\varepsilon_m &= \frac{1}{2}\left(f_+ - f_- \right),\\
		f_\pm &= \sqrt{\delta^2 + 2t^2\left[1 \pm \cos\left(\varphi_x/2\right)\right]},\label{equation:fpm}
	\end{align}
\end{subequations}
where $\delta = E_\text{tot} + \varepsilon$ is the energy penalty for moving charge from the island to the dot. 
For small $t/\delta$, the second term in~\cref{equation:majorana-box-diagonal-0} moreover reduces to~\cref{equation:majorana-transmon-tunneling}, since
\begin{equation}
	\varepsilon_m \simeq \frac{t^2}{\delta} \cos\left(\frac{\varphi_x}{2}\right).
\end{equation}
In the opposite limit, if the chemical potential of the quantum dot is tuned such that $\varepsilon = -E_\text{tot}$ and the energy penalty to occupy the dot $\delta=0$, then $\varepsilon_m \sim t$.

\subsection{Dispersive interaction with a resonator}\label{subsection:majorana-box-dispersive}

As with the Majorana transmon, we can read out the logical state by coupling the qubit to a resonator. 
There are essentially two options for engineering a dipole coupling by capacitively coupling to the resonator. Either the resonator voltage can be (predominantly) coupled to the dot or (predominantly) to the superconducting island. The key requirement for readout is that the resonator must be sensitive to the movement of charge between the superconducting island and the dot, and the two coupling schemes are in that sense equivalent (as shown, e.g., in Ref.~\cite{grimsmo2019majorana}).
The two choices might, however, have different practical advantages and disadvantages; in particular, stronger coupling may be possible by coupling to the island.
We here focus mainly on capacitive coupling to the superconducting island charge, for concreteness, but we emphasize that our results apply equally well to both schemes.
The qubit-resonator interaction is thus, in analogy with~\cref{equation:majorana-transmon-resonator-interaction}, given by
\begin{equation}\label{equation:majorana-box-resonator-interaction}
	\hat{H}_\inter = i \hbar \lambda \hat N_\text{tot} (\hat a\dg - \hat a).
\end{equation}

We perform the same unitary transformation that led to~\cref{equation:majorana-box-diagonal-0}, and again set $t_{L,R}=t$, $n_g=0$ and project onto the subspace with zero overall charge, to find an interaction in this subspace of the form (see~\cref{appendix:majorana-box-diagonalization})
\begin{equation}\label{equation:majorana-box-resonator-interaction-diagonal}
	\begin{aligned}
		\hat{H}_{\inter,0} ={}&i\hbar\left[g_c\hat{c}^\dagger\hat{c} + \frac{g_m}{2}(i\hat{\gamma}_2\hat{\gamma}_3 +1) \right](\hat{a}^\dagger-\hat{a})
		\\ &+ \frac{i\hbar}{2}\left[g_+(\hat{\gamma}_2+i\hat{\gamma}_3)\hat{c} + \mathrm{H.c.}\right](\hat{a}^\dagger-\hat{a}),
		\\ &+ \frac{i\hbar}{2}\left[g_-(\hat{\gamma}_2-i\hat{\gamma}_3)\hat{c} + \mathrm{H.c.}\right](\hat{a}^\dagger-\hat{a}),
	\end{aligned}
\end{equation}
where
\begin{subequations}\label{equation:majorana-box-resonator-interaction-diagonal-coefficients}
    \begin{align}
        g_c &= -\frac{\lambda}{2}\left(\frac{\delta}{f_+}+\frac{\delta}{f_-}\right),
        \\ g_m &= -\frac{\lambda}{2}\left(\frac{\delta}{f_+}-\frac{\delta}{f_-}\right),
        \\ g_\pm &= -\frac{\lambda}{2} \frac{i t(e^{\frac{i\varphi_x}{2}}\pm 1)}{f_\pm}\label{equation:gpm}.
    \end{align}
\end{subequations}
We note that, in the case where the quantum dot is resonant with the island and $\delta = 0$, then $g_c = g_m = 0$ and $|g_\pm| = \lambda/2$.
For the alternative choice of coupling the resonator to the dot, simply replace $\hat N_\text{tot} \to \hat d^\dagger \hat d$ in~\cref{equation:majorana-box-resonator-interaction} and the above results still apply with a sign change $\lambda\to-\lambda$ in~\cref{equation:majorana-box-resonator-interaction-diagonal-coefficients}~\cite{grimsmo2019majorana}.

From the second and third lines of~\cref{equation:majorana-box-resonator-interaction-diagonal} we see that, in this frame, the resonator induces a transition that involves moving an electron from the dot to the island and flipping the Majorana parity $i \hat\gamma_2 \hat\gamma_3$. 
The energy difference corresponding to this transition is $\varepsilon_c \pm \varepsilon_m= f_\pm$, depending on the state of the Majorana degree of freedom.

\begin{figure}
	\includegraphics[trim=0 0 0 0, width=\linewidth]{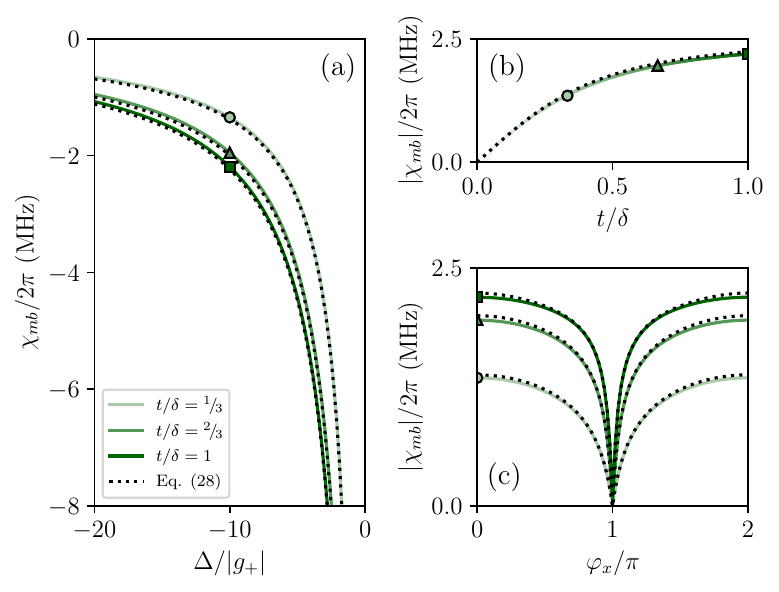}
	\caption{
	   Dispersive frequency shifts of a readout resonator coupled to a Majorana box qubit $\chi_\mb$ (green) for $\delta/h = (E_\mathrm{tot} + \varepsilon)/h = 5$ GHz, $n_g=0$, $t_{L,R} = t$, and $\lambda/2\pi = 100$ MHz. 
	   In (a), (b) $\varphi_x = 0$ and in (b), (c) $\Delta/g_+=-10$. 
	   Circle, triangle, and square markers indicate equivalent points between plots. 
	   We show $\chi_\mb$ as a function of (a) detuning $\Delta = f_+/\hbar-\omega_\res$ for three values of $t/\delta$,
	   (b) $t/\delta$ for $\Delta/|g_+|=-10$, (c) $\varphi_x=0$.
	   Analytical results (dotted) obtained from~\cref{equation:majorana-box-dispersive-shift} are also shown. 
	   A very small discrepancy arises due to the rotating- wave approximation used in~\cref{equation:majorana-box-dispersive-shift}.
	    }
	\label{figure:majorana-box-dispersive-shift}
\end{figure}

Having diagonalized the Majorana box qubit Hamiltonian, it is straight-forward to use the second-order Schrieffer-Wolff formula~\cref{equation:qubit-dispersive-shift} to obtain an analytical expression for the qubit-state-dependent dispersive shift. 
Under a rotating wave approximation for the resonator-qubit interaction, we find that
\begin{equation}\label{equation:majorana-box-dispersive-shift}
    \chi_\mb \simeq \frac{1}{2}\left(\frac{|g_+|^2}{f_+/\hbar-\omega_\res} - \frac{|g_-|^2}{f_-/\hbar-\omega_\res}\right),
\end{equation}
where we assume that $\varepsilon_c > \varepsilon_m$.

\begin{figure*}
    \includegraphics[width=\linewidth]{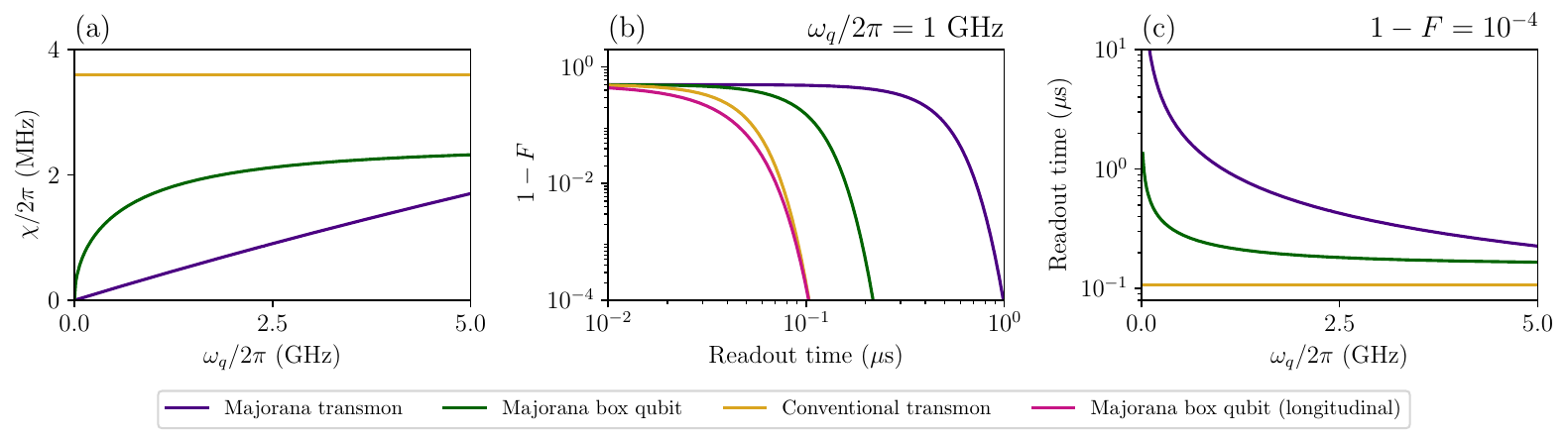}
    \caption{ 
        Timescales and fidelities for resonator-based readout of Majorana qubits. 
        (a) Dispersive shift of a readout resonator coupled to a Majorana transmon qubit $\chi_\mt$ (purple), a Majorana box qubit $\chi_\mb$ (green), and a conventional transmon $\chi_t$ (gold) as a function of the qubit frequency $\omega_q/2\pi$ for $n_g=0$, $\varphi_x=0$, $\lambda/2\pi = 100$ MHz, and fixed detuning $\Delta/g = -10$. 
        For the Majorana and conventional transmon we have set $E_C/h = 250$ MHz and $E_J/E_C = 50$. 
        For the Majorana box qubit, we have set $\delta/ h = (E_\mathrm{tot} + \varepsilon)/h = 5$ GHz.
        (b) Infidelity $1-F$ of each readout scheme as a function of readout time $\tau$, for $\omega_q/2\pi = 1$ GHz, $\bar n /n_\mathrm{crit} = 1/5$, and $\kappa = 2\chi$. 
        Longitudinal case calculated for $\tilde g_z/2\pi = 10$ MHz, as detailed in~\cref{appendix:majorana-box-longitudinal}.
        (c) Readout time required to reach a measurement fidelity of 99.99\% for each dispersive scheme in (a) as a function of $\omega_q/2\pi$.
        }
    \label{figure:readout-estimates}
\end{figure*}

We compare~\cref{equation:majorana-box-dispersive-shift} to a numerical diagonalization of the qubit Hamiltonian, following the same procedure as for the Majorana transmon qubit, to extract the qubit-dependent dispersive shift $\chi_\mb$. 
In~\cref{figure:majorana-box-dispersive-shift}~(a) we show $\chi_\mb$ as a function of $\Delta/|g_+|$ for different tunneling strengths $t_{L,R}=t$, where $\Delta \equiv f_+/\hbar - \omega_\res$. 
As the Majorana modes hybridize with the dot, the qubit splitting and the dispersive shifts grow larger, also shown in~\cref{figure:majorana-box-spectra}~(b). 
As with the Majorana transmon, the dispersive shift $\chi_\mb$ depends on the offset flux $\varphi_x$, as shown in~\cref{figure:majorana-box-dispersive-shift}~(c).
Again, care must be taken to ensure that $\varphi_x \neq \pi$.
However, in contrast to the Majorana transmon, the dependence is more favorable, leading to a wider range of $\varphi_x$ where $\chi_\mb$ is close to its maximum possible magnitude.

Our results show that the energy scale of the dispersive shifts for the Majorana box qubit are comparable to the Majorana transmon qubit for comparable ratios between resonator coupling strengths and detuning $\Delta/g$. 
A more detailed comparison is made in~\cref{section:readout-estimates}.

\section{Readout times and fidelities}\label{section:readout-estimates}

In this section we calculate estimates of the timescales and fidelities of the Majorana qubit readout schemes presented in the previous sections. 
The key results are presented in~\cref{figure:readout-estimates}.
It is important to note that these estimates have been obtained for an idealized situation where the dispersive approximation is assumed to be valid, and no noise or decoherence is included beyond the dephasing caused by the measurement itself. 
These results are therefore not meant to be quantitative predictions for the measurement fidelity in an experiment, but serve to compare the speed and fidelity for different qubit types: the conventional transmon, Majorana transmon and the Majorana box qubit. For the Majorana box qubit, we also compare dispersive readout to longitudinal readout; see~\cref{appendix:majorana-box-longitudinal}.

Our predictions of the dispersive shifts of the Majorana transmon and Majorana box qubit as functions of qubit frequency $\omega_q$ are compared in~\cref{figure:readout-estimates}~(a).
To make this comparison, we fix the value of $\Delta/g = -10$, where $g \in \{g_\tl, g_+\}$ [see~\cref{equation:transmon-resonator-coupling,equation:gpm}] and $\Delta \in \{\omega_+ - \omega_r, f_+/\hbar - \omega_r\}$ [see~\cref{equation:majorana-transmon-transition-splittings,equation:fpm}] for the Majorana transmon and the Majorana box qubit, respectively. 
We also include the dispersive shift of a conventional transmon, for which we use $g=g_\tl$ and $\Delta = \omega_t-\omega_r$.
In other words, $g$ and $\Delta$ quantify the relevant coupling strength and resonator detuning for each qubit type, respectively.

We observe that, for smaller qubit frequencies (corresponding to weaker MZM interaction energies), the Majorana box qubit produces larger dispersive shifts than the Majorana transmon for the same value of $\Delta/g$.
Nevertheless, both variants may achieve dispersive shifts in the megahertz regime for reasonable parameters, comparable to conventional transmon qubits~\cite{walter2017rapid} and the recent demonstration of a nanowire quantum dot readout in Ref.~\cite{de2019rapid}.

The qubit-state-dependent phase shift that arises during dispersive coupling allows for readout of the qubit by probing the resonator at its resonant frequency. 
The size of the dispersive shift $\chi$ directly determines the rate at which this phase shift can be resolved to a given fidelity.  
We quantify this effect with the signal-to-noise ratio (SNR) for a heterodyne measurement of the resonator output field. 
To simplify the treatment, we consider an idealized situation with unit efficiency measurement and no additional noise or decoherence, such that the qubit-dependent response of the resonator is Gaussian. 
In this case an analytical form for the SNR can be found~\cite{didier2015fast, blais2020circuit}:
\begin{equation}\label{snr}
    \mathrm{SNR} = 2 |\epsilon| \sqrt{\frac{\tau}{\chi}} \left[ 1 - \frac{1}{\chi\tau} \left( 1 - e^{-\chi\tau}\cos{\chi\tau}\right) \right].
\end{equation}
Here $|\epsilon|$ is the amplitude of the resonator drive and $\tau$ is the measurement time.  
The SNR will in general depend on the resonator damping rate $\kappa$, and we have set $\kappa=2\chi$ to give the optimal SNR at long integration times~\cite{blais2020circuit,didier2015fast}. 

The assignment fidelity can be related to the SNR through~\cite{grimsmo2019majorana,blais2020circuit}
\begin{equation}\label{equation:fidelity}
    F = 1 - \half\left[p(1|0) + p(0|1) \right]
    = 1 - \half \mathrm{erfc}\left(\frac{\mathrm{SNR}}{2}\right),
\end{equation}
where $p(i|j)$ is the probability of obtaining the outcome $i$ given that the qubit was in the state $\ket{j}$.
(We note that an alternative definition of fidelity $F_m=1 - p(1|0) - p(0|1) = 2F-1$ is also often used~\cite{blais2020circuit}.)
We emphasize that these results hold for the ideal, dispersive Hamiltonian~\cref{equation:dispersive-interaction}. In other words, the dispersive approximation is assumed to be valid.
This approximation will break down for large photon numbers~\cite{blais2020circuit}.

From these expressions, we calculate the expected measurement infidelities $1-F$ for each qubit as a function of integration time $\tau$ at $\omega_q/2\pi = 1$ GHz in \cref{figure:readout-estimates}~(b).  
We have chosen the resonator drive strength $\epsilon$ such that $\bar n/n_\mathrm{crit} = 1/5$, where $\bar n=2(\epsilon/\kappa)^2$ is the resonator photon number and $n_\mathrm{crit}\equiv(\Delta/2g)^2$. The latter can be thought of as a rough measure of when the dispersive approximation is expected to break down~\cite{blais2020circuit}.

For comparison, we also show the infidelity of a longitudinal readout scheme for the Majorana box qubit in~\cref{figure:readout-estimates}~(b).
We have chosen parameters such that $\kappa$ and $\bar n$ are equal between the dispersive and longitudinal cases, which for these parameters correspond to a modulation of the longitudinal coupling strength by $\tilde g_z/2\pi \simeq 10$ MHz. 
As shown in~\cref{figure:majorana-box-longitudinal} in~\cref{appendix:majorana-box-longitudinal} such a modulation can be achieved by a very modest modulation in either the tunnel coupling or external flux.
It is noteworthy that longitudinal readout gives a much faster (and thus higher fidelity) readout for this modest value of parametric modulation.
For example, doubling the modulation amplitude translates to a readout that is roughly twice as fast.

Finally, we calculate the measurement integration time required to achieve a measurement fidelity of $99.99\%$ for the dispersive readout protocols as a function of qubit splitting $\omega_q$, shown in \cref{figure:readout-estimates}~(d). 
For the chosen system parameters and assumptions, both Majorana qubits may achieve high-fidelity dispersive measurements in a fraction of a microsecond. 
Furthermore, the Majorana box qubit, which produces a larger dispersive shift, benefits from a faster readout time at the same value of $\Delta/g$ and qubit frequency.

\section{Conclusions}\label{section:conclusions}

Our results are very promising for dispersive readout as a means to measure Majorana qubits quickly and with high fidelity.
We have calculated the qubit-dependent dispersive shifts of a readout resonator for Majorana transmons and Majorana box qubits, under a simple capacitive coupling of the resonator to the qubit. 
This dispersive shift can be used to readout the state of the qubit by measuring the phase shift of a resonant probe tone on the resonator. 
We find that the dispersive shift for Majorana qubits of both types can be in the megahertz range for reasonable parameters.
These results are encouraging, as they indicate that well-established and extremely successful readout techniques can be adopted from the circuit QED context~\cite{walter2017rapid, blais2020circuit}.

There are some key differences in the QND nature of dispersive readout for a Majorana transmon compared to a Majorana box qubit.
For the Majorana transmon, the qubit-resonator interaction manifestly preserves the Majorana parity, independent of the detuning of the readout resonator from the relevant transitions between qubit energy levels.
This protection originates from the fact that both $\hat H_\mtmon$ in~\cref{equation:majorana-transmon} and $\hat H_\inter$ in~\cref{equation:majorana-transmon-resonator-interaction} commute with $i \hat\gamma_2 \hat\gamma_3$. 
The Majorana parity is therefore preserved independently of whether the perturbative dispersive approximation $H_\disp$,~\cref{equation:dispersive-interaction}, is valid.
As a result, the dispersive readout of a Majorana transmon is quantum nondemolition in a stronger sense than for conventional charge qubits.

For the Majorana box qubit, the situation is different.
Here, coupling to the resonator is induced by tunneling of charge from the qubit island to a nearby quantum dot.
In a readout scheme, the tunnel coupling should be turned on adiabatically, such that the system evolves into dressed joint eigenstates of the qubit-dot system [see~\cref{equation:majorana-box-diagonal}]. 
However, the interaction with the resonator induces transitions between dressed eigenstates of different Majorana parity; see~\cref{equation:majorana-box-resonator-interaction-diagonal}. 
A quantum nondemolition readout is therefore only approximately recovered in a limit where the relevant transition frequency for moving an electron between the island and the dot is far detuned from the resonator frequency, leading to $\hat H_\disp$ in~\cref{equation:dispersive-interaction}. 
The readout is therefore no longer QND when the dispersive approximation breaks down, which can happen, e.g., for large photon numbers.
It is worth noting, however, that the \emph{joint} Majorana-dot parity is manifestly conserved by~\cref{equation:majorana-box-resonator-interaction-diagonal}.
If one can also determine the state of the dot, prior to \emph{and} after the measurement, with high fidelity, it may be possible to confirm the Majorana parity, as has been discussed in Refs.~\cite{steiner2020readout,munk2020paritytocharge,khindanov2020visibility}.

The fundamental differences between the Majorana transmon and the Majorana box qubits makes a quantitative comparison of readout fidelity and speed more challenging. In~\cref{section:readout-estimates} we compared the two qubits for equal qubit energy splitting, and at a fixed value of coupling strength relative to resonator detuning, $g/\Delta$, and fixed value of resonator photons relative to $n_\text{crit} \equiv (\Delta/2g)^2$.
With this choice, our results suggest that the Majorana box qubit produces larger dispersive shifts, and may therefore enjoy a faster readout.
However, because the breakdown of the dispersive interaction will manifest itself differently for the two qubits, this comparison might not be fair. 
In particular, the performance of dispersive readout for large photon numbers requires further study.

The two qubits will also likely have differing robustness to noise.  To achieve high-fidelity readout, it is crucial that the readout time is much smaller than the coherence times of the qubit.
The Majorana box qubit is expected to be more robust against quasiparticle poisoning due to the use of smaller superconducting islands~\cite{karzig2020quasiparticle}. 
Both qubits will suffer dephasing due to finite nanowire lengths~\cite{knapp2018dephasing}. 
Charge noise might reduce the readout SNR due to fluctuations in system parameters, thus leading to a longer readout time (see, e.g., Refs~\cite{maman2020charge,khindanov2020visibility} for a more detailed discussion).

We also draw attention to the functional dependence of the dispersive shift on flux that threads the relevant loops for each respective qubit, shown in~\cref{figure:majorana-transmon-dispersive-shift}~(d) and~\cref{figure:majorana-box-dispersive-shift}~(c). 
With the likelihood that offset fluxes are present in the system, and given some distribution of these between different qubits, a challenge for the scalability of the (measurement-based) approaches is the requirement to locally tune the flux for each qubit in order to maximize readout fidelity.
From this perspective we find that the Majorana box qubits are favourable, since flux tuning is likely only necessary for a limited number of qubits that have offsets very close to $\varphi_x = \pi$.

Finally, we note that the longitudinal readout protocol introduced in Ref.~\cite{grimsmo2019majorana} is entirely independent of the resonator detuning, and can therefore be used in a regime where the parity breaking terms for the Majorana box qubit are negligible (corresponding to a regime where the dispersive shift is negligible).
Our results moreover show that longitudinal readout may lead to even faster and higher fidelity readout in practice, given that a reasonable parametric modulation is possible.

\begin{acknowledgments}
	We thank Andrew Doherty and Torsten Karzig for useful discussions. 
	This research was supported by the Australian Research Council, through the Centre of Excellence for Engineered Quantum Systems (EQUS) project number CE170100009 and Discovery Early Career Research Award project number DE190100380.
\end{acknowledgments}

\appendix

\section{Light-matter interaction in the dispersive regime}\label{appendix:dispersive-regime-derivation}

We here review the general theory of dispersive coupling between an arbitrary multilevel system, which we here simply refer to as an ``artificial atom,'' and a resonator. 
We treat the resonator as a single mode, for simplicity. 
In this section we provide a high-level summary of Ref.~\cite{zhu2013circuit}, and we refer the reader to that work for further details. 
We use this theory to calculate dispersive shifts for the two variants of Majorana qubits in~\cref{section:majorana-transmon,section:majorana-box}.

Consider a generic artificial atom with eigenstates $\ket{l}$ and eigenenergies $\hbar \omega_l$, described by a Hamiltonian $\hat H_\atom = \sum_l \hbar \omega_l \ket l \bra l$. 
We assume that the artificial atom is coupled to a resonator described in a single-mode approximation by a Hamiltonian $\hat H_\res = \hbar \omega_\res \hat a\dg \hat a$. 
The atom-resonator coupling has the form $\hat H_\inter = i \hbar \lambda \hat N (\hat a\dg - \hat a)$, where $\hat N$ is the number operator for the artificial atom charge degree of freedom, which couples to the voltage of the resonator $\hat V_r \sim i(\hat a\dg -  \hat a)$ with interaction strength $\lambda$. 
Expressing the interaction in the eigenbasis of the artificial atom, the two coupled systems are described by a Hamiltonian
\begin{equation}\label{equation:atom-resonator-interaction}
    \hat H = \hbar \omega_\res \hat a\dg \hat a + \sum_l \hbar\omega_l |l\rangle\langle l | \\
    + \sum_{l,l^\prime} \hbar g_{l,l^\prime} | l \rangle \langle l^\prime |  \left( \hat a\dg - \hat a \right). 
\end{equation}
The coefficients $g_{l,l^\prime} = i \lambda \langle l | \hat N | l^\prime \rangle$ are matrix elements for transitions between states in the multilevel artificial atom. 
The dispersive regime refers to a situation where the frequency detunings between all relevant atomic transitions and the resonator, $\Delta_{l,l^\prime} = \omega_l - \omega_{l^\prime} - \omega_\res$, are large relative to their respective transition amplitudes $g_{l,l^\prime}$.
Energy exchange between the artificial atom and the resonator becomes a virtual process and we can write an effective Hamiltonian for~\cref{equation:atom-resonator-interaction} that is derived from second-order Schrieffer-Wolff perturbation theory~\cite{zhu2013circuit, bravyi2011schrieffer, winkler2003spin}
\begin{equation}\label{equation:atom-resonator-dispersive-interaction}
    \begin{aligned}
        \hat{H}_\disp ={}& \hbar\omega_\res \hat a\dg \hat a + \sum_l \hbar(\omega_l + \eta_l) | l \rangle\langle l | + \sum_l \hbar\chi_l \hat a\dg \hat a | l \rangle\langle l |.
    \end{aligned}
\end{equation}
At this level of approximation the interaction is diagonal both with respect to the artificial atom and resonator eigenstates. 
Here, $\chi_l$ is a state-dependent frequency shift to the resonator, and $\eta_l$ is a correction to $\omega_l$, given by
\begin{subequations}\label{equation:atom-resonator-dispersive-interaction-coefficients}
    \begin{align}
        \chi_l &= \sum_{l^\prime} \chi_{l,l^\prime} -  \chi_{l^\prime, l},\label{equation:atom-resonator-dispersive-shift-l} \\
        \eta_l &= \sum_{l^\prime} \chi_{l,l^\prime},\\
        \chi_{l,l^\prime} &= \frac{|g_{l,l^\prime}|^2}{\Delta_{l,l^\prime}}.
    \end{align}
\end{subequations}
If a qubit is encoded in two states $\{\ket{l=0}, \ket{l=1}\}$ of the artificial atom then we obtain \cref{equation:dispersive-interaction} with $\hat\sigma_z = \ket{1}\bra{1} - \ket{0}\bra{0}$, $\omega_q = \omega_1 - \omega_0 + \eta_1 - \eta_0$, and 
\begin{equation}\label{equation:qubit-dispersive-shift}
    \chi_q = (\chi_1 - \chi_0)/2.
\end{equation}

In this paper, we analyze the qubit-state-dependent dispersive frequency shift $\chi_q$ for $q \in \{\tl, \mt, \mb\}$, corresponding to the conventional transmon qubit, the Majorana transmon qubit, and the Majorana box qubit, respectively.
The choice of logical qubit states $\ket{l=0}$ and $\ket{l=1}$ are different for each qubit: $\{\ket g, \ket e\}$ for the conventional transmon, $\{\ket{g,+},\ket{g,-}\}$ for the Majorana transmon, and for the Majorana box qubit the two dressed states $\{\overline{\ket{N,n_d,+}}, \overline{\ket{N,n_d,-}}\}$ adiabatically connected to the two degenerate ground states when the tunnel couplings are zero, $t_{L,R}=0$. 
We denote the corresponding dispersive shifts by
\begin{subequations}\label{equation:qubit-dipserive-shift-list}
    \begin{align}
        \chi_\tl &= (\chi_e - \chi_g)/2, \\
        \chi_\mt &= (\chi_{g,-} - \chi_{g,+})/2, \\
        \chi_\mb &= (\chi_{N,n_d,+} - \chi_{N,n_d,-})/2.
    \end{align}
\end{subequations}

\section{Modeling interactions between MZMs}\label{appendix:majorana-interaction-models}

\begin{figure}
    \includegraphics[trim=0 0 0 0, width=\linewidth]{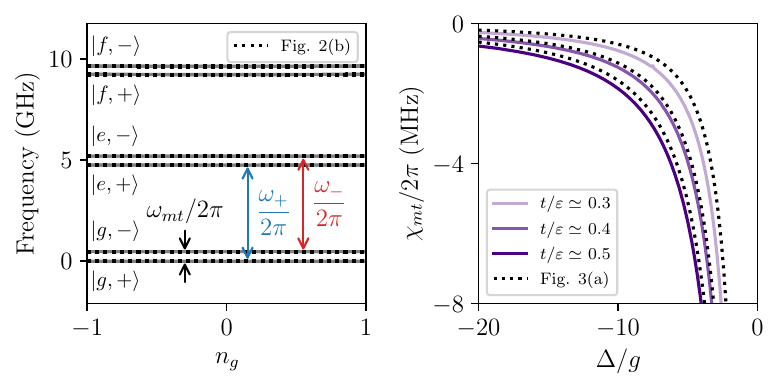}
    \caption{
        Low-energy spectrum and dispersive shifts of a Majorana transmon qubit with an ``indirect'' interaction model. System parameters are identical to those in \cref{figure:majorana-transmon-dispersive-shift}(a). 
        Here $\varepsilon/h = 20$ GHz and $t_{L,R}=t$ is tuned such that the energy splitting between the two lowest levels, $\hbar\omega_{mt}$, is equal to the corresponding case from~\cref{figure:majorana-transmon-dispersive-shift}(a).
        }
    \label{figure:majorana-transmon-tunneling-model-comparison}
\end{figure}

In this section we demonstrate that the ``direct'' interaction term,~\cref{equation:majorana-transmon-tunneling}, for a Majorana transmon is equivalent to a model where the interaction is mediated by a quantum dot, in the limit where the energy cost for moving an electron onto the dot is sufficiently large. 
Consider a pair of topological superconductors, such as those shown in~\cref{figure:majorana-qubits}~(a). 
Adjacent MZMs $\hat\gamma_2$ and $\hat\gamma_3$ are separated by an insulating-semiconducting region. 
Interaction between the Majorana modes is controlled by a gate voltage, which tunes the chemical potential in the semiconducting region. 
When the chemical potential is sufficiently large, the region forms a potential barrier that prohibits interaction between the MZMs. 
However, when the height of the potential barrier is lowered, electronic bound states in the semiconducting region become energetically accessible.
For simplicity, we consider a single energy level, described by fermionic operator $\hat d$ and Hamiltonian $\hat H_d = \varepsilon \hat d\dg \hat d$.
The MZMs couple to the dot, described by a tunneling Hamiltonian~\cite{fu2010electron}
\begin{equation}
    \hat H_\tunn = \frac{1}{2} \left[\left( it_L e^{\frac{i \hat\varphi_L}{2}} \hat\gamma_2 - t_R e^{\frac{i \hat\varphi_R}{2}} \hat\gamma_ 3 \right) \hat d + \text{H.c.} \right].
\end{equation}
This ``indirect'' model, wherein the MZMs interact via virtual occupation of the quantum dot can be compared to a ``direct'' interaction, \cref{equation:majorana-transmon-tunneling}~\cite{hell2016time}:
\begin{equation}
    \hat{H}_\maj = -{i} E_M \hat\gamma_2 \hat\gamma_3 \cos{ \left( \frac{\hat\varphi + \varphi_x}{2} \right) },
\end{equation}
with $\hat\varphi = \hat\varphi_L - \hat\varphi_R$.
The two models agree when $\delta = \varepsilon + E_C$ is large relative to the tunnel couplings $t_{L,R}$, where $E_C$ is the charging energy due to capacitive coupling between the two topological superconductors, from~\cref{equation:transmon}.
To demonstrate this, we have numerically plotted the spectrum and dispersive shifts of a Majorana transmon qubit for both interaction terms in~\cref{figure:majorana-transmon-tunneling-model-comparison}.

\section{Dispersive shift for the Majorana transmon qubit}\label{appendix:majorana-transmon-schrieffer-wolff}

In~\cref{subsection:majorana-transmon-dispersive-regime} we provided an approximate formula for the dispersive frequency shift of a resonator coupled to a Majorana transmon qubit,~\cref{equation:majorana-transmon-dispersive-shift-effective}.
These were obtained analytically by applying the Schrieffer-Wolff method detailed in~\cref{appendix:dispersive-regime-derivation} to the complete Majorana transmon qubit-resonator Hamiltonian
\begin{equation}
    \hat H = \hat H_\res + \hat H_\mtmon + \hat H_\inter.
\end{equation}
This treatment requires the qubit Hamiltonian 
\begin{equation}
    \hat H_\mtmon = \hat H_\tmon + \hat H_\maj,
\end{equation}
to be written in a diagonal form.
We use an approximation, where $\hat H_\tmon$ and $\hat H_\maj$ are given in~\cref{equation:transmon-effective} and~\cref{equation:majorana-transmon-tunneling-effective}, respectively.
The qubit-resonator interaction Hamiltonian has the specified form
\begin{equation}
    \hat H_\inter = i \hbar \lambda \hat N (\hat a\dg - \hat a),
\end{equation}
where we use $\hat N = {i} \left({E_J}/{2E_C}\right)^{1/4} \big(\hat b\dg - \hat b\big)$. 
We proceed by calculating the dispersive frequency shift of the resonator
\begin{equation}\label{equation:majorana-transmon-dispersive-shift-definition}
    \chi_\mt = (\chi_{g,-} - \chi_{g,+})/2,
\end{equation}
for the logical qubit states $\ket 0 = \ket{g,+}$ and $\ket 1 = \ket{g,-}$.
The only nonzero matrix elements for this effective model are 
\begin{subequations}
    \begin{align}
        |g_{j,j+1}| &= |i\lambda\braket{j,a|\hat N|j+1,a}\!| = g_\tl\sqrt{j+1},
        \\  |g_{j,j-1}| &= |i\lambda\braket{j,a|\hat N|j-1,a}\!| = g_\tl\sqrt{j},
    \end{align}
\end{subequations}
where $g_\tl = \lambda({E_J}/{2E_C})^{1/4}$.
Specifically, in the logical subspace, we have $\ket{g,+} \leftrightarrow \ket{e,+}$ with frequency $\omega_+$, and $\ket{g,-} \leftrightarrow \ket{e,-}$ with frequency $\omega_-$, as indicated in~\cref{figure:majorana-transmon-spectra}. 
From~\cref{equation:atom-resonator-dispersive-shift-l}, we obtain
\begin{equation}
    \begin{aligned}
        \chi_{g,+} &\simeq \frac{| i \lambda \braket{g,+|\hat N|e,+} |^2}{-\omega_+ - \omega_\res}-\frac{| i \lambda \braket{e,+|\hat N|g,+} |^2}{\omega_+ - \omega_\res}, \\
         &= -\frac{g_\tl^2}{\omega_+ + \omega_\res} - \frac{g_\tl^2}{\omega_+ - \omega_\res},
    \end{aligned}
\end{equation}
and
\begin{equation}
    \begin{aligned}
        \chi_{g,-} &\simeq \frac{| i \lambda \braket{g,-|\hat N|e,-} |^2}{-\omega_- - \omega_\res} - \frac{| i \lambda \braket{e,-|\hat N|g,-} |^2}{\omega_- - \omega_\res}, \\
        &= -\frac{g_\tl^2}{\omega_- + \omega_\res} - \frac{g_\tl^2}{\omega_- - \omega_\res}.
    \end{aligned}
\end{equation}
Finally, from~\cref{equation:majorana-transmon-dispersive-shift-definition} we have, 
\begin{equation}
    \chi_\mt \simeq \frac{1}{2} \left(\frac{g_\tl^2}{\omega_+-\omega_\res} - \frac{g_\tl^2}{\omega_--\omega_\res}\right).
\end{equation}
where we have dropped fast-rotating terms approximately $1/(\omega_\pm + \omega_r)$ for simplicity. 

\section{Dispersive shift for the Majorana box qubit}\label{appendix:majorana-box-schrieffer-wolff}

\subsection{Diagonalizing the Majorana box qubit Hamiltonian}\label{appendix:majorana-box-diagonalization}

The complete Hamiltonian for the Majorana box qubit can be written as
\begin{equation}\label{equation:majorana-box-jordan-wigner}
    \begin{aligned}
        \hat H_\mbx ={}&
        E_\text{tot} (\hat N_\text{tot} - n_g)^2
        + \varepsilon \hat\sigma_d\dg \hat\sigma_d\\
        +\frac{1}{2}\Big[ e^{\frac{i \hat\varphi}{2}} &\left( it_L e^{i \varphi_x/2} \hat X_m \hat\sigma_d + t_R \hat Y_m \hat\sigma_d \right) + \text{H.c.} \Big].
    \end{aligned}
\end{equation}
We have performed a Jordan-Wigner transformation
\begin{equation}\label{equation:jordan-wigner}
    \begin{aligned}
        \hat d &= -\hat\sigma_d,\\
        \hat\gamma_2 &= - \hat Z_d \hat X_m,\\
        \hat\gamma_3 &= - \hat Z_d \hat Y_m,
    \end{aligned}
\end{equation}
where $\hat\sigma_j = \ket 0_j \bra 1$ is a two-level system lowering operator, and $\hat X_j = \hat\sigma_j\dg + \hat\sigma_j$, $\hat Y_j = i(\hat\sigma_j\dg - \hat\sigma_j)$ and $\hat Z_j = \hat\sigma_j \hat\sigma_j\dg - \hat\sigma_j\dg \hat \sigma_j$ are the usual Pauli matrices with $j=d,m$.

To diagonalize~\cref{equation:majorana-box-jordan-wigner}, we follow Ref.~\cite{grimsmo2019majorana}. 
We can label a basis for the system by $\ket{N,n_d,a}$, where $N\in\mathbb Z$, $n_d=0,1$ and $a=\pm$ correspond to the island charge, dot occupation number and Majorana parity, respectively. 
Note that the total charge $\hat N + \hat\sigma_d\dg \hat\sigma_d$ is conserved by~\cref{equation:majorana-box-jordan-wigner}.
We therefore define projectors 
\begin{subequations}
    \begin{align}
        \hat P_n ={}& \ket{0_n} \bra{0_n} + \ket{1_n} \bra{1_n},\\
        \ket {0_n} ={}& \ket{N=n,n_d=0},\\
        \ket {1_n} ={}& \ket{N=n-1,n_d=1},
    \end{align}
\end{subequations}
and use the fact that $\hat H_\mbx = \sum_n \hat H_{\mbx,n}$ with 
\begin{equation}
    \hat H_{\mbx,n} = \hat P_n \hat H_\mbx \hat P_n.
\end{equation}

The Hamiltonian projected onto the $n$th subspace can be diagonalized by a unitary (see Ref.~\cite{grimsmo2019majorana} for details)
\begin{equation}
    \hat U_n = e^{-\left[\alpha_-(n) \hat\sigma_n\dg \hat \sigma_m - \alpha_+(n) \hat \sigma_n\dg \hat \sigma_m\dg - \hc \right]},
\end{equation}
with
\begin{equation}
    \tan\left[2\left|\alpha_\pm(n)\right|\right] = \frac{|t_L e^{i\phi_x/2} \pm t_R|}{2\delta(n)},
\end{equation}
and we have defined a lowering operator within the $n$th subspace
\begin{equation}
    \hat\sigma_n = \hat P_n e^{\frac{i\hat\varphi}{2}} \hat\sigma_d \hat P_n = \ket{0_n}\bra{1_n}.
\end{equation}
Explicitly, we find that
\begin{equation}\label{equation:majorana-box-diagonal-n}
    \begin{aligned}
        \hat H_{\mbx,n}' &= \hat U_n\dg \hat H_{\mbx,n} \hat U_n \\
        &=\varepsilon_c(n) \hat{\sigma}_n^\dagger \hat{\sigma}_n
        +\varepsilon_m(n)\hat\sigma_m\dg \hat \sigma_m + E(n)\\
    \end{aligned}
\end{equation}
where 
\begin{subequations}\label{equation:majorana-box-diagonal-coefficients}
    \begin{align}
        \varepsilon_c(n) &= \frac{\text{sgn}\delta(n)}{2}\left[ f_+(n)+f_-(n) \right],
        \\ \varepsilon_m(n) &= \frac{\text{sgn}\delta(n)}{2}\left[ f_+(n)-f_-(n) \right],
        \\ E(n) &= E_\mathrm{tot} (n-n_g)^2 + \frac{\delta(n) - \varepsilon_c(n) - \varepsilon_m(n)}{2},
        \\ f_\pm (n) &= \sqrt{\delta(n)^2 + t_L^2 + t_R^2  \pm 2 t_L t_R \cos\left(\frac{\varphi_x}{2}\right)},
        \\ \delta(n) &= E_\mathrm{tot} + \varepsilon - 2E_\mathrm{tot}(n-n_g).
    \end{align}
\end{subequations}
To recover the full Hamiltonian given in~\cref{equation:majorana-box-diagonal}, we use $\hat H'_\mbx = \sum_n \hat H'_{\mbx, n}\hat P_n$ and reverse the Jordan-Wigner transformation, \cref{equation:jordan-wigner}.

The qubit-resonator interaction Hamiltonian $\hat H_\inter$, defined in~\cref{equation:majorana-box-resonator-interaction}, also preserves the total charge and can similarly be written as $\hat H_\inter = \sum_n \hat H_{\inter,n} \hat P_n$ with
\begin{equation}
    \hat H_{\inter,n} =  i\hbar\lambda\left(n - \hat \sigma_n\dg \hat \sigma_n \right)(\hat a\dg - \hat a).
\end{equation}
Following Ref.~\cite{grimsmo2019majorana}, we perform the same unitary transformation as in~\cref{equation:majorana-box-diagonal-n} to find that
\begin{equation}
    \begin{aligned}
        \hat H_{\inter,n}' ={}& \hat U_n\dg \hat H_{\inter,n} \hat U_n = i\hbar\lambda \hat{N}_n' (\hat{a}^\dagger - \hat{a}),
    \end{aligned}
\end{equation}
where
\begin{equation}
    \begin{aligned}\label{equation:majorana-box-resonator-charge-operator}
        \lambda\hat N'_n ={}&\lambda n + g_c(n)\hat{\sigma}_n^\dagger\hat{\sigma}_n + g_m(n)\hat\sigma_m\dg\hat\sigma_m
        \\ &+ \left[g_+(n) \hat\sigma_m \hat{\sigma}_n + g_-(n) \hat\sigma_m\dg \hat{\sigma}_n + \mathrm{H.c.}\right],
    \end{aligned}
\end{equation}
and
\begin{subequations}
    \begin{align}
        g_c(n) &= -\frac{\lambda}{2}\left(\frac{\delta(n)}{f_+(n)}+\frac{\delta(n)}{f_-(n)}\right),
        \\ g_m(n) &= -\frac{\lambda}{2}\left(\frac{\delta(n)}{f_+(n)}-\frac{\delta(n)}{f_-(n)}\right),
        \\ g_\pm(n) &= -\frac{\lambda}{2} \frac{i(t_Le^{\frac{i\varphi_x}{2}}\pm t_R)}{f_\pm(n)}.
    \end{align}
\end{subequations}
The result in~\cref{equation:majorana-box-diagonal-0} is found by setting $n_g=0$, $t_{L,R}=t$, projecting onto the $n=0$ total charge sector, and reversing the Jordan-Wigner transformation.

\subsection{Calculating the dispersive shift}

We can now calculate the dispersive shift from the complete qubit-resonator Hamiltonian in the $n$th subspace
\begin{equation}
    \hat H'_n = \hat H_\res + \hat H'_{\mbx,n} + \hat H'_{\inter, n}.   
\end{equation}

The eigenstates of $\hat H_{\mbx,n}'$ for each subspace $n$ can be labeled $\ket{n-\sigma_n,\sigma_n, a}$ with $\sigma_n=0,1$ the occupancy of $\hat \sigma_n\dg \hat \sigma_n$ and $a=\pm$ corresponding to the Majorana parity $2\hat \sigma_m\dg \hat \sigma_m - 1$. 
We proceed by calculating the dispersive frequency shift of the resonator
\begin{equation}\label{equation:majorana-box-dispersive-shift-definition}
    \chi_{\mb,n} = (\chi_{n,0,+} - \chi_{n,0,-})/2,
\end{equation}
for the two states $\ket{1}=\ket{n,0,+}$ and $\ket{0}=\ket{n,0,-}$, where we assume that the dot is unoccupied for the two qubit states.
From~\cref{equation:majorana-box-resonator-charge-operator} we have the following nonzero, off-diagonal matrix elements:
\begin{subequations}
    \begin{align}
        |i\lambda\braket{n,0,-| \hat N_n'|n-1,1,+}| &= |g_+(n)|,
        \\ |i\lambda\braket{n,0,+| \hat N_n'|n-1,1,-}| &= |g_-(n)|.
    \end{align}
\end{subequations}
The diagonal matrix elements, corresponding to the first three terms in~\cref{equation:majorana-box-resonator-charge-operator}, cancel out when calculating the dispersive shifts by~\cref{equation:atom-resonator-dispersive-shift-l}.

In the logical qubit subspace we have two transitions: $\ket{n,0,+} \leftrightarrow \ket{n-1,1,-}$ with energy $\varepsilon_c(n)-\varepsilon_m(n) = f_-(n)$ and $\ket{n,0,-} \leftrightarrow \ket{n-1,1,+}$ with energy $\varepsilon_c(n)+\varepsilon_m(n) = f_+(n)$.
From~\cref{equation:atom-resonator-dispersive-shift-l}, we obtain
\begin{equation}
    \begin{aligned}
        \chi_{n,0,+} ={}& \frac{| i \lambda \braket{n,0,+|\hat N_n'|n-1,1,-} |^2}{-f_-(n)/\hbar - \omega_\res}
        \\ &-\frac{| i \lambda \braket{n-1,1,-|\hat N_n'|n,0,+} |^2}{f_-(n)/\hbar - \omega_\res} 
        \\ ={}& -\frac{|g_-(n)|^2}{f_-(n)/\hbar + \omega_\res} - \frac{|g_-(n)|^2}{f_-(n)/\hbar  - \omega_\res},
    \end{aligned}
\end{equation}
and
\begin{equation}
    \begin{aligned}
        \chi_{n,0,-} ={}& \frac{| i \lambda \braket{n,0,-|\hat N_n'|n-1,1,+} |^2}{-f_+(n)/\hbar - \omega_\res}
        \\ &-\frac{| i \lambda \braket{n-1,1,+|\hat N_n'|n,0,-} |^2}{f_+(n)/\hbar - \omega_\res} 
        \\ ={}& -\frac{|g_+(n)|^2}{f_+(n)/\hbar + \omega_\res} - \frac{|g_+(n)|^2}{f_+(n)/\hbar  - \omega_\res}.
    \end{aligned}
\end{equation}
Finally, from~\cref{equation:majorana-box-dispersive-shift-definition} we have, 
\begin{equation}
    \chi_{\mb,n} \simeq \frac{1}{2}\left(\frac{|g_+(n)|^2}{f_+(n)/\hbar-\omega_\res} - \frac{|g_-(n)|^2}{f_-(n)/\hbar-\omega_\res}\right),
\end{equation}
where we have again dropped fast-rotating terms approximately $1/(f_\pm(n)/\hbar + \omega_r)$ for simplicity. 
The result quoted in~\cref{equation:majorana-box-dispersive-shift} corresponds to setting $n=0$, $n_g=0$ and $t_{L,R}=t$.

\section{Longitudinal readout for Majorana box qubits}\label{appendix:majorana-box-longitudinal}

As alluded to in~\cref{section:light-matter}, the parity protection of the Majorana box qubit dispersive readout scheme is not as strong as that of the Majorana transmon.
This is apparent from the last two lines of~\cref{equation:majorana-box-resonator-interaction-diagonal}, since the qubit-resonator interaction does not commute with the Majorana parity operator $i\hat\gamma_2\hat\gamma_3$ in the diagonal frame of $\hat H_\mbx'$. 
Approximate parity protection is recovered in the dispersive regime where~\cref{equation:dispersive-interaction} is valid.

\begin{figure}
    \centering
    \includegraphics[width=\linewidth]{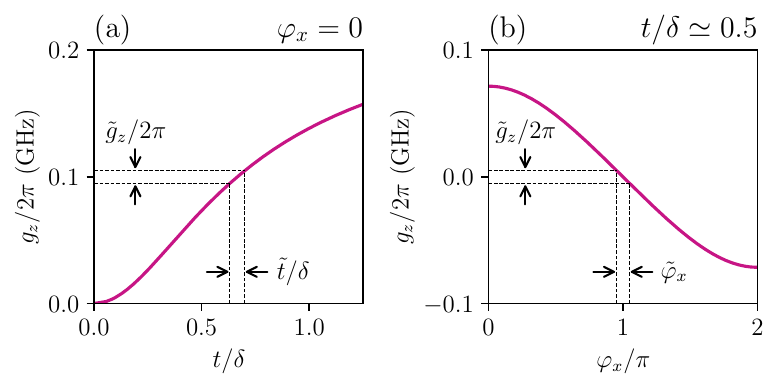}
    \caption{
        Longitudinal coupling strength for the Majorana box qubit $g_z=g_m/2$ as a function of $t/\delta$ and $\varphi_x$. 
        For (a), we have set $\varphi_x = 0$ and, for (b), we have set $t/\delta \simeq 0.5$.
        Dashed lines indicate a modulation of $\tilde g_z = 10$ MHz, which was used in~\cref{figure:readout-estimates}.
        This corresponds to a modulation of roughly (a) $\tilde t/\delta \simeq 0.1$ or (b) $\tilde \varphi_x = \pi/10$.
        }
    \label{figure:majorana-box-longitudinal}
\end{figure}

An alternative approach was proposed in Ref.~\cite{grimsmo2019majorana} that instead relies on exploiting the \emph{longitudinal} Majorana-resonator interaction apparent in the first line of~\cref{equation:majorana-box-resonator-interaction-diagonal}, quantified by $g_m$.
Longitudinal coupling refers to an interaction that is diagonal in the qubit basis, in contrast to the off-diagonal, or transversal, coupling usually exploited for charge qubits. 
The readout scheme introduced in Ref.~\cite{grimsmo2019majorana} uses a parametric modulation of the longitudinal coupling $g_m$ in~\cref{equation:majorana-box-resonator-interaction-diagonal}. 
In practice, this can be achieved in a number of ways, including flux modulation and modulating the tunnel coupling $t$, and/or the detuning $\delta$.

The first line of~\cref{equation:majorana-box-resonator-interaction-diagonal} clearly commutes with $i\hat\gamma_2\hat\gamma_3$, and a QND readout protocol can therefore be achieved given that the parity nonconserving processes in the two last lines are heavily suppressed. 
This latter requirement can be met by placing the resonator sufficiently far off-resonance from the relevant island-dot transitions. 
More precisely, for ideal longitudinal readout, the denominators $|f_\pm/\hbar-\omega_r|$ in~\cref{equation:majorana-box-dispersive-shift} should be sufficiently large such that $\chi_\mb$ is negligible. 
This should be relatively straight forward to achieve, e.g., by increasing the charging and dot energies, $E_\text{tot}$ and $\varepsilon$. 
We note that increasing these energy scales might furthermore be beneficial from the point of view of quasiparticle poisoning~\cite{plugge2017majorana,karzig2017scalable}.

The infidelity of a longitudinal readout scheme for the Majorana box qubit is shown in~\cref{figure:readout-estimates}~(b). When producing these results, we take the coupling strength to be modulated at the resonator frequency, $g_m(t) = \bar g_m + \tilde g_m \cos(\omega_r t)$, and assume an ideal longitudinal interaction, which in a frame rotating at $\omega_r$ can be written as
\begin{equation}
\hat H_z = \frac{i \tilde g_z}{2} i \hat \gamma_2 \hat \gamma_3 (\hat a^\dagger - \hat a),
\end{equation}
where $\tilde g_z = \tilde g_m/2$ is the longitudinal modulation and fast rotating terms are neglected.

We refer the reader to Refs.~\cite{didier2015fast,grimsmo2019majorana} for further details.
Here the fidelity is calculated using~\cref{equation:fidelity} with the following expression for the SNR~\cite{didier2015fast}:
\begin{equation}
    \mathrm{SNR}_l = \sqrt{8} | \tilde g_z | \sqrt{\frac{\tau}{\kappa}}  \left[ 1 - \frac{2}{\kappa\tau} \left( 1 - e^{-\frac12\kappa\tau}\right) \right].
\end{equation}

For the sake of comparison in~\cref{figure:readout-estimates}, we set $\kappa$ and the intracavity photon number $\bar n = (\tilde g_z/\kappa)^2$~\cite{didier2015fast} to be the same for the dispersive and longitudinal readout protocols.
For the parameters in the figure, this corresponds to a modulation of $\tilde g_z/2\pi \simeq 10$ MHz.
For concreteness, we illustrate the range of modulation of $t/\delta$ and $\varphi_x$, respectively, needed to achieve $\tilde g_z/2\pi = 10$ MHz in~\cref{figure:majorana-box-longitudinal} for an example parameter set. 
These results clearly show that even a modest modulation in external parameters can lead to extremely fast longitudinal readout.


\begin{thebibliography}{60}%
\makeatletter
\providecommand \@ifxundefined [1]{%
 \@ifx{#1\undefined}
}%
\providecommand \@ifnum [1]{%
 \ifnum #1\expandafter \@firstoftwo
 \else \expandafter \@secondoftwo
 \fi
}%
\providecommand \@ifx [1]{%
 \ifx #1\expandafter \@firstoftwo
 \else \expandafter \@secondoftwo
 \fi
}%
\providecommand \natexlab [1]{#1}%
\providecommand \enquote  [1]{``#1''}%
\providecommand \bibnamefont  [1]{#1}%
\providecommand \bibfnamefont [1]{#1}%
\providecommand \citenamefont [1]{#1}%
\providecommand \href@noop [0]{\@secondoftwo}%
\providecommand \href [0]{\begingroup \@sanitize@url \@href}%
\providecommand \@href[1]{\@@startlink{#1}\@@href}%
\providecommand \@@href[1]{\endgroup#1\@@endlink}%
\providecommand \@sanitize@url [0]{\catcode `\\12\catcode `\$12\catcode
  `\&12\catcode `\#12\catcode `\^12\catcode `\_12\catcode `\%12\relax}%
\providecommand \@@startlink[1]{}%
\providecommand \@@endlink[0]{}%
\providecommand \url  [0]{\begingroup\@sanitize@url \@url }%
\providecommand \@url [1]{\endgroup\@href {#1}{\urlprefix }}%
\providecommand \urlprefix  [0]{URL }%
\providecommand \Eprint [0]{\href }%
\providecommand \doibase [0]{https://doi.org/}%
\providecommand \selectlanguage [0]{\@gobble}%
\providecommand \bibinfo  [0]{\@secondoftwo}%
\providecommand \bibfield  [0]{\@secondoftwo}%
\providecommand \translation [1]{[#1]}%
\providecommand \BibitemOpen [0]{}%
\providecommand \bibitemStop [0]{}%
\providecommand \bibitemNoStop [0]{.\EOS\space}%
\providecommand \EOS [0]{\spacefactor3000\relax}%
\providecommand \BibitemShut  [1]{\csname bibitem#1\endcsname}%
\let\auto@bib@innerbib\@empty
\bibitem [{\citenamefont {Kitaev}(2003)}]{kitaev2003fault}%
  \BibitemOpen
  \bibfield  {author} {\bibinfo {author} {\bibfnamefont {A.~Y.}\ \bibnamefont
  {Kitaev}},\ }\bibfield  {title} {\bibinfo {title} {{Fault-tolerant quantum
  computation by anyons}},\ }\href
  {https://doi.org/10.1016/s0003-4916(02)00018-0} {\bibfield  {journal}
  {\bibinfo  {journal} {Ann. Phys.}\ }\textbf {\bibinfo {volume} {303}},\
  \bibinfo {pages} {2–30} (\bibinfo {year} {2003})}\BibitemShut {NoStop}%
\bibitem [{\citenamefont {Bravyi}(2006)}]{bravyi2006universal}%
  \BibitemOpen
  \bibfield  {author} {\bibinfo {author} {\bibfnamefont {S.}~\bibnamefont
  {Bravyi}},\ }\bibfield  {title} {\bibinfo {title} {{Universal quantum
  computation with the $\nu=5/2$ fractional quantum Hall state}},\ }\href
  {https://doi.org/10.1103/PhysRevA.73.042313} {\bibfield  {journal} {\bibinfo
  {journal} {Phys. Rev. A}\ }\textbf {\bibinfo {volume} {73}},\ \bibinfo
  {pages} {042313} (\bibinfo {year} {2006})}\BibitemShut {NoStop}%
\bibitem [{\citenamefont {Nayak}\ \emph {et~al.}(2008)\citenamefont {Nayak},
  \citenamefont {Simon}, \citenamefont {Stern}, \citenamefont {Freedman},\ and\
  \citenamefont {Das~Sarma}}]{nayak2008non}%
  \BibitemOpen
  \bibfield  {author} {\bibinfo {author} {\bibfnamefont {C.}~\bibnamefont
  {Nayak}}, \bibinfo {author} {\bibfnamefont {S.~H.}\ \bibnamefont {Simon}},
  \bibinfo {author} {\bibfnamefont {A.}~\bibnamefont {Stern}}, \bibinfo
  {author} {\bibfnamefont {M.}~\bibnamefont {Freedman}},\ and\ \bibinfo
  {author} {\bibfnamefont {S.}~\bibnamefont {Das~Sarma}},\ }\bibfield  {title}
  {\bibinfo {title} {{Non-Abelian anyons and topological quantum
  computation}},\ }\href {https://doi.org/10.1103/RevModPhys.80.1083}
  {\bibfield  {journal} {\bibinfo  {journal} {Rev. Mod. Phys.}\ }\textbf
  {\bibinfo {volume} {80}},\ \bibinfo {pages} {1083} (\bibinfo {year}
  {2008})}\BibitemShut {NoStop}%
\bibitem [{\citenamefont {Kitaev}(2001)}]{kitaev2001unpaired}%
  \BibitemOpen
  \bibfield  {author} {\bibinfo {author} {\bibfnamefont {A.~Y.}\ \bibnamefont
  {Kitaev}},\ }\bibfield  {title} {\bibinfo {title} {{Unpaired Majorana
  fermions in quantum wires}},\ }\href
  {https://doi.org/10.1070/1063-7869/44/10s/s29} {\bibfield  {journal}
  {\bibinfo  {journal} {Phys.-Usp.}\ }\textbf {\bibinfo {volume} {44}},\
  \bibinfo {pages} {131–136} (\bibinfo {year} {2001})}\BibitemShut {NoStop}%
\bibitem [{\citenamefont {Fu}\ and\ \citenamefont
  {Kane}(2008)}]{fu2008superconducting}%
  \BibitemOpen
  \bibfield  {author} {\bibinfo {author} {\bibfnamefont {L.}~\bibnamefont
  {Fu}}\ and\ \bibinfo {author} {\bibfnamefont {C.~L.}\ \bibnamefont {Kane}},\
  }\bibfield  {title} {\bibinfo {title} {{Superconducting Proximity Effect and
  Majorana Fermions at the Surface of a Topological Insulator}},\ }\href
  {https://doi.org/10.1103/PhysRevLett.100.096407} {\bibfield  {journal}
  {\bibinfo  {journal} {Phys. Rev. Lett.}\ }\textbf {\bibinfo {volume} {100}},\
  \bibinfo {pages} {096407} (\bibinfo {year} {2008})}\BibitemShut {NoStop}%
\bibitem [{\citenamefont {Lutchyn}\ \emph {et~al.}(2010)\citenamefont
  {Lutchyn}, \citenamefont {Sau},\ and\ \citenamefont
  {Das~Sarma}}]{lutchyn2010majorana}%
  \BibitemOpen
  \bibfield  {author} {\bibinfo {author} {\bibfnamefont {R.~M.}\ \bibnamefont
  {Lutchyn}}, \bibinfo {author} {\bibfnamefont {J.~D.}\ \bibnamefont {Sau}},\
  and\ \bibinfo {author} {\bibfnamefont {S.}~\bibnamefont {Das~Sarma}},\
  }\bibfield  {title} {\bibinfo {title} {{Majorana Fermions and a Topological
  Phase Transition in Semiconductor-Superconductor Heterostructures}},\ }\href
  {https://doi.org/10.1103/PhysRevLett.105.077001} {\bibfield  {journal}
  {\bibinfo  {journal} {Phys. Rev. Lett.}\ }\textbf {\bibinfo {volume} {105}},\
  \bibinfo {pages} {077001} (\bibinfo {year} {2010})}\BibitemShut {NoStop}%
\bibitem [{\citenamefont {Oreg}\ \emph {et~al.}(2010)\citenamefont {Oreg},
  \citenamefont {Refael},\ and\ \citenamefont {von Oppen}}]{oreg2010helical}%
  \BibitemOpen
  \bibfield  {author} {\bibinfo {author} {\bibfnamefont {Y.}~\bibnamefont
  {Oreg}}, \bibinfo {author} {\bibfnamefont {G.}~\bibnamefont {Refael}},\ and\
  \bibinfo {author} {\bibfnamefont {F.}~\bibnamefont {von Oppen}},\ }\bibfield
  {title} {\bibinfo {title} {{Helical Liquids and Majorana Bound States in
  Quantum Wires}},\ }\href {https://doi.org/10.1103/PhysRevLett.105.177002}
  {\bibfield  {journal} {\bibinfo  {journal} {Phys. Rev. Lett.}\ }\textbf
  {\bibinfo {volume} {105}},\ \bibinfo {pages} {177002} (\bibinfo {year}
  {2010})}\BibitemShut {NoStop}%
\bibitem [{\citenamefont {Sau}\ \emph {et~al.}(2010)\citenamefont {Sau},
  \citenamefont {Lutchyn}, \citenamefont {Tewari},\ and\ \citenamefont
  {Das~Sarma}}]{sau2010generic}%
  \BibitemOpen
  \bibfield  {author} {\bibinfo {author} {\bibfnamefont {J.~D.}\ \bibnamefont
  {Sau}}, \bibinfo {author} {\bibfnamefont {R.~M.}\ \bibnamefont {Lutchyn}},
  \bibinfo {author} {\bibfnamefont {S.}~\bibnamefont {Tewari}},\ and\ \bibinfo
  {author} {\bibfnamefont {S.}~\bibnamefont {Das~Sarma}},\ }\bibfield  {title}
  {\bibinfo {title} {{Generic New Platform for Topological Quantum Computation
  Using Semiconductor Heterostructures}},\ }\href
  {https://doi.org/10.1103/PhysRevLett.104.040502} {\bibfield  {journal}
  {\bibinfo  {journal} {Phys. Rev. Lett.}\ }\textbf {\bibinfo {volume} {104}},\
  \bibinfo {pages} {040502} (\bibinfo {year} {2010})}\BibitemShut {NoStop}%
\bibitem [{\citenamefont {Alicea}(2012)}]{alicea2012new}%
  \BibitemOpen
  \bibfield  {author} {\bibinfo {author} {\bibfnamefont {J.}~\bibnamefont
  {Alicea}},\ }\bibfield  {title} {\bibinfo {title} {{New directions in the
  pursuit of Majorana fermions in solid state systems}},\ }\href
  {https://doi.org/10.1088/0034-4885/75/7/076501} {\bibfield  {journal}
  {\bibinfo  {journal} {Rep. Prog. Phys.}\ }\textbf {\bibinfo {volume} {75}},\
  \bibinfo {pages} {076501} (\bibinfo {year} {2012})}\BibitemShut {NoStop}%
\bibitem [{\citenamefont {Mourik}\ \emph {et~al.}(2012)\citenamefont {Mourik},
  \citenamefont {Zuo}, \citenamefont {Frolov}, \citenamefont {Plissard},
  \citenamefont {Bakkers},\ and\ \citenamefont
  {Kouwenhoven}}]{mourik2012signatures}%
  \BibitemOpen
  \bibfield  {author} {\bibinfo {author} {\bibfnamefont {V.}~\bibnamefont
  {Mourik}}, \bibinfo {author} {\bibfnamefont {K.}~\bibnamefont {Zuo}},
  \bibinfo {author} {\bibfnamefont {S.~M.}\ \bibnamefont {Frolov}}, \bibinfo
  {author} {\bibfnamefont {S.~R.}\ \bibnamefont {Plissard}}, \bibinfo {author}
  {\bibfnamefont {E.~P. A.~M.}\ \bibnamefont {Bakkers}},\ and\ \bibinfo
  {author} {\bibfnamefont {L.~P.}\ \bibnamefont {Kouwenhoven}},\ }\bibfield
  {title} {\bibinfo {title} {{Signatures of Majorana Fermions in Hybrid
  Superconductor-Semiconductor Nanowire Devices}},\ }\href
  {https://doi.org/10.1126/science.1222360} {\bibfield  {journal} {\bibinfo
  {journal} {Science}\ }\textbf {\bibinfo {volume} {336}},\ \bibinfo {pages}
  {1003–1007} (\bibinfo {year} {2012})}\BibitemShut {NoStop}%
\bibitem [{\citenamefont {Beenakker}(2013)}]{beenakker2013search}%
  \BibitemOpen
  \bibfield  {author} {\bibinfo {author} {\bibfnamefont {C.~W.~J.}\
  \bibnamefont {Beenakker}},\ }\bibfield  {title} {\bibinfo {title} {{Search
  for Majorana Fermions in Superconductors}},\ }\href
  {https://doi.org/10.1146/annurev-conmatphys-030212-184337} {\bibfield
  {journal} {\bibinfo  {journal} {Annu. Rev. Condens. Matter Phys.}\ }\textbf
  {\bibinfo {volume} {4}},\ \bibinfo {pages} {113} (\bibinfo {year}
  {2013})}\BibitemShut {NoStop}%
\bibitem [{\citenamefont {Nadj-Perge}\ \emph {et~al.}(2014)\citenamefont
  {Nadj-Perge}, \citenamefont {Drozdov}, \citenamefont {Li}, \citenamefont
  {Chen}, \citenamefont {Jeon}, \citenamefont {Seo}, \citenamefont {MacDonald},
  \citenamefont {Bernevig},\ and\ \citenamefont
  {Yazdani}}]{nadj2014observation}%
  \BibitemOpen
  \bibfield  {author} {\bibinfo {author} {\bibfnamefont {S.}~\bibnamefont
  {Nadj-Perge}}, \bibinfo {author} {\bibfnamefont {I.~K.}\ \bibnamefont
  {Drozdov}}, \bibinfo {author} {\bibfnamefont {J.}~\bibnamefont {Li}},
  \bibinfo {author} {\bibfnamefont {H.}~\bibnamefont {Chen}}, \bibinfo {author}
  {\bibfnamefont {S.}~\bibnamefont {Jeon}}, \bibinfo {author} {\bibfnamefont
  {J.}~\bibnamefont {Seo}}, \bibinfo {author} {\bibfnamefont {A.~H.}\
  \bibnamefont {MacDonald}}, \bibinfo {author} {\bibfnamefont {B.~A.}\
  \bibnamefont {Bernevig}},\ and\ \bibinfo {author} {\bibfnamefont
  {A.}~\bibnamefont {Yazdani}},\ }\bibfield  {title} {\bibinfo {title}
  {{Observation of Majorana fermions in ferromagnetic atomic chains on a
  superconductor}},\ }\href {https://doi.org/10.1126/science.1259327}
  {\bibfield  {journal} {\bibinfo  {journal} {Science}\ }\textbf {\bibinfo
  {volume} {346}},\ \bibinfo {pages} {602} (\bibinfo {year}
  {2014})}\BibitemShut {NoStop}%
\bibitem [{\citenamefont {Albrecht}\ \emph {et~al.}(2016)\citenamefont
  {Albrecht}, \citenamefont {Higginbotham}, \citenamefont {Madsen},
  \citenamefont {Kuemmeth}, \citenamefont {Jespersen}, \citenamefont {Nygård},
  \citenamefont {Krogstrup},\ and\ \citenamefont
  {Marcus}}]{albrecht2016exponential}%
  \BibitemOpen
  \bibfield  {author} {\bibinfo {author} {\bibfnamefont {S.~M.}\ \bibnamefont
  {Albrecht}}, \bibinfo {author} {\bibfnamefont {A.~P.}\ \bibnamefont
  {Higginbotham}}, \bibinfo {author} {\bibfnamefont {M.}~\bibnamefont
  {Madsen}}, \bibinfo {author} {\bibfnamefont {F.}~\bibnamefont {Kuemmeth}},
  \bibinfo {author} {\bibfnamefont {T.~S.}\ \bibnamefont {Jespersen}}, \bibinfo
  {author} {\bibfnamefont {J.}~\bibnamefont {Nygård}}, \bibinfo {author}
  {\bibfnamefont {P.}~\bibnamefont {Krogstrup}},\ and\ \bibinfo {author}
  {\bibfnamefont {C.~M.}\ \bibnamefont {Marcus}},\ }\bibfield  {title}
  {\bibinfo {title} {{Exponential protection of zero modes in Majorana
  islands}},\ }\href {https://doi.org/10.1038/nature17162} {\bibfield
  {journal} {\bibinfo  {journal} {Nature}\ }\textbf {\bibinfo {volume} {531}},\
  \bibinfo {pages} {206–209} (\bibinfo {year} {2016})}\BibitemShut {NoStop}%
\bibitem [{\citenamefont {Deng}\ \emph {et~al.}(2016)\citenamefont {Deng},
  \citenamefont {Vaitiekėnas}, \citenamefont {Hansen}, \citenamefont {Danon},
  \citenamefont {Leijnse}, \citenamefont {Flensberg}, \citenamefont {Nygård},
  \citenamefont {Krogstrup},\ and\ \citenamefont {Marcus}}]{deng2016majorana}%
  \BibitemOpen
  \bibfield  {author} {\bibinfo {author} {\bibfnamefont {M.~T.}\ \bibnamefont
  {Deng}}, \bibinfo {author} {\bibfnamefont {S.}~\bibnamefont {Vaitiekėnas}},
  \bibinfo {author} {\bibfnamefont {E.~B.}\ \bibnamefont {Hansen}}, \bibinfo
  {author} {\bibfnamefont {J.}~\bibnamefont {Danon}}, \bibinfo {author}
  {\bibfnamefont {M.}~\bibnamefont {Leijnse}}, \bibinfo {author} {\bibfnamefont
  {K.}~\bibnamefont {Flensberg}}, \bibinfo {author} {\bibfnamefont
  {J.}~\bibnamefont {Nygård}}, \bibinfo {author} {\bibfnamefont
  {P.}~\bibnamefont {Krogstrup}},\ and\ \bibinfo {author} {\bibfnamefont
  {C.~M.}\ \bibnamefont {Marcus}},\ }\bibfield  {title} {\bibinfo {title}
  {{Majorana bound state in a coupled quantum-dot hybrid-nanowire system}},\
  }\href {https://doi.org/10.1126/science.aaf3961} {\bibfield  {journal}
  {\bibinfo  {journal} {Science}\ }\textbf {\bibinfo {volume} {354}},\ \bibinfo
  {pages} {1557–1562} (\bibinfo {year} {2016})}\BibitemShut {NoStop}%
\bibitem [{\citenamefont {Nichele}\ \emph {et~al.}(2017)\citenamefont
  {Nichele}, \citenamefont {Drachmann}, \citenamefont {Whiticar}, \citenamefont
  {O'Farrell}, \citenamefont {Suominen}, \citenamefont {Fornieri},
  \citenamefont {Wang}, \citenamefont {Gardner}, \citenamefont {Thomas},
  \citenamefont {Hatke}, \citenamefont {Krogstrup}, \citenamefont {Manfra},
  \citenamefont {Flensberg},\ and\ \citenamefont
  {Marcus}}]{nichele2017scaling}%
  \BibitemOpen
  \bibfield  {author} {\bibinfo {author} {\bibfnamefont {F.}~\bibnamefont
  {Nichele}}, \bibinfo {author} {\bibfnamefont {A.~C.~C.}\ \bibnamefont
  {Drachmann}}, \bibinfo {author} {\bibfnamefont {A.~M.}\ \bibnamefont
  {Whiticar}}, \bibinfo {author} {\bibfnamefont {E.~C.~T.}\ \bibnamefont
  {O'Farrell}}, \bibinfo {author} {\bibfnamefont {H.~J.}\ \bibnamefont
  {Suominen}}, \bibinfo {author} {\bibfnamefont {A.}~\bibnamefont {Fornieri}},
  \bibinfo {author} {\bibfnamefont {T.}~\bibnamefont {Wang}}, \bibinfo {author}
  {\bibfnamefont {G.~C.}\ \bibnamefont {Gardner}}, \bibinfo {author}
  {\bibfnamefont {C.}~\bibnamefont {Thomas}}, \bibinfo {author} {\bibfnamefont
  {A.~T.}\ \bibnamefont {Hatke}}, \bibinfo {author} {\bibfnamefont
  {P.}~\bibnamefont {Krogstrup}}, \bibinfo {author} {\bibfnamefont {M.~J.}\
  \bibnamefont {Manfra}}, \bibinfo {author} {\bibfnamefont {K.}~\bibnamefont
  {Flensberg}},\ and\ \bibinfo {author} {\bibfnamefont {C.~M.}\ \bibnamefont
  {Marcus}},\ }\bibfield  {title} {\bibinfo {title} {{Scaling of Majorana
  Zero-Bias Conductance Peaks}},\ }\href
  {https://doi.org/10.1103/PhysRevLett.119.136803} {\bibfield  {journal}
  {\bibinfo  {journal} {Phys. Rev. Lett.}\ }\textbf {\bibinfo {volume} {119}},\
  \bibinfo {pages} {136803} (\bibinfo {year} {2017})}\BibitemShut {NoStop}%
\bibitem [{\citenamefont {Lutchyn}\ \emph {et~al.}(2018)\citenamefont
  {Lutchyn}, \citenamefont {Bakkers}, \citenamefont {Kouwenhoven},
  \citenamefont {Krogstrup}, \citenamefont {Marcus},\ and\ \citenamefont
  {Oreg}}]{lutchyn2018majorana}%
  \BibitemOpen
  \bibfield  {author} {\bibinfo {author} {\bibfnamefont {R.~M.}\ \bibnamefont
  {Lutchyn}}, \bibinfo {author} {\bibfnamefont {E.~P. A.~M.}\ \bibnamefont
  {Bakkers}}, \bibinfo {author} {\bibfnamefont {L.~P.}\ \bibnamefont
  {Kouwenhoven}}, \bibinfo {author} {\bibfnamefont {P.}~\bibnamefont
  {Krogstrup}}, \bibinfo {author} {\bibfnamefont {C.~M.}\ \bibnamefont
  {Marcus}},\ and\ \bibinfo {author} {\bibfnamefont {Y.}~\bibnamefont {Oreg}},\
  }\bibfield  {title} {\bibinfo {title} {{Majorana zero modes in
  superconductor–semiconductor heterostructures}},\ }\href
  {https://doi.org/10.1038/s41578-018-0003-1} {\bibfield  {journal} {\bibinfo
  {journal} {Nat. Rev. Mater.}\ }\textbf {\bibinfo {volume} {3}},\ \bibinfo
  {pages} {52–68} (\bibinfo {year} {2018})}\BibitemShut {NoStop}%
\bibitem [{\citenamefont {Leijnse}\ and\ \citenamefont
  {Flensberg}(2012)}]{leijnse2012introduction}%
  \BibitemOpen
  \bibfield  {author} {\bibinfo {author} {\bibfnamefont {M.}~\bibnamefont
  {Leijnse}}\ and\ \bibinfo {author} {\bibfnamefont {K.}~\bibnamefont
  {Flensberg}},\ }\bibfield  {title} {\bibinfo {title} {{Introduction to
  topological superconductivity and Majorana fermions}},\ }\href
  {https://doi.org/10.1088/0268-1242/27/12/124003} {\bibfield  {journal}
  {\bibinfo  {journal} {Semicond. Sci. Technol.}\ }\textbf {\bibinfo {volume}
  {27}},\ \bibinfo {pages} {124003} (\bibinfo {year} {2012})}\BibitemShut
  {NoStop}%
\bibitem [{\citenamefont {Stanescu}\ and\ \citenamefont
  {Tewari}(2013)}]{stanescu2013majorana}%
  \BibitemOpen
  \bibfield  {author} {\bibinfo {author} {\bibfnamefont {T.~D.}\ \bibnamefont
  {Stanescu}}\ and\ \bibinfo {author} {\bibfnamefont {S.}~\bibnamefont
  {Tewari}},\ }\bibfield  {title} {\bibinfo {title} {{Majorana fermions in
  semiconductor nanowires: fundamentals, modeling, and experiment}},\ }\href
  {https://doi.org/10.1088/0953-8984/25/23/233201} {\bibfield  {journal}
  {\bibinfo  {journal} {J. Phys. Condens. Matter}\ }\textbf {\bibinfo {volume}
  {25}},\ \bibinfo {pages} {233201} (\bibinfo {year} {2013})}\BibitemShut
  {NoStop}%
\bibitem [{\citenamefont {Alicea}\ \emph {et~al.}(2011)\citenamefont {Alicea},
  \citenamefont {Oreg}, \citenamefont {Refael}, \citenamefont {von Oppen},\
  and\ \citenamefont {Fisher}}]{alicea2011non}%
  \BibitemOpen
  \bibfield  {author} {\bibinfo {author} {\bibfnamefont {J.}~\bibnamefont
  {Alicea}}, \bibinfo {author} {\bibfnamefont {Y.}~\bibnamefont {Oreg}},
  \bibinfo {author} {\bibfnamefont {G.}~\bibnamefont {Refael}}, \bibinfo
  {author} {\bibfnamefont {F.}~\bibnamefont {von Oppen}},\ and\ \bibinfo
  {author} {\bibfnamefont {M.~P.~A.}\ \bibnamefont {Fisher}},\ }\bibfield
  {title} {\bibinfo {title} {{Non-Abelian statistics and topological quantum
  information processing in 1D wire networks}},\ }\href
  {https://doi.org/https://doi.org/10.1038/nphys1915} {\bibfield  {journal}
  {\bibinfo  {journal} {Nat. Phys.}\ }\textbf {\bibinfo {volume} {7}},\
  \bibinfo {pages} {412} (\bibinfo {year} {2011})}\BibitemShut {NoStop}%
\bibitem [{\citenamefont {Sau}\ \emph {et~al.}(2011)\citenamefont {Sau},
  \citenamefont {Clarke},\ and\ \citenamefont {Tewari}}]{sau2011controlling}%
  \BibitemOpen
  \bibfield  {author} {\bibinfo {author} {\bibfnamefont {J.~D.}\ \bibnamefont
  {Sau}}, \bibinfo {author} {\bibfnamefont {D.~J.}\ \bibnamefont {Clarke}},\
  and\ \bibinfo {author} {\bibfnamefont {S.}~\bibnamefont {Tewari}},\
  }\bibfield  {title} {\bibinfo {title} {{Controlling non-Abelian statistics of
  Majorana fermions in semiconductor nanowires}},\ }\href
  {https://doi.org/10.1103/PhysRevB.84.094505} {\bibfield  {journal} {\bibinfo
  {journal} {Phys. Rev. B}\ }\textbf {\bibinfo {volume} {84}},\ \bibinfo
  {pages} {094505} (\bibinfo {year} {2011})}\BibitemShut {NoStop}%
\bibitem [{\citenamefont {van Heck}\ \emph {et~al.}(2012)\citenamefont {van
  Heck}, \citenamefont {Akhmerov}, \citenamefont {Hassler}, \citenamefont
  {Burrello},\ and\ \citenamefont {Beenakker}}]{van2012coulomb}%
  \BibitemOpen
  \bibfield  {author} {\bibinfo {author} {\bibfnamefont {B.}~\bibnamefont {van
  Heck}}, \bibinfo {author} {\bibfnamefont {A.~R.}\ \bibnamefont {Akhmerov}},
  \bibinfo {author} {\bibfnamefont {F.}~\bibnamefont {Hassler}}, \bibinfo
  {author} {\bibfnamefont {M.}~\bibnamefont {Burrello}},\ and\ \bibinfo
  {author} {\bibfnamefont {C.~W.~J.}\ \bibnamefont {Beenakker}},\ }\bibfield
  {title} {\bibinfo {title} {{Coulomb-assisted braiding of Majorana fermions in
  a Josephson junction array}},\ }\href
  {https://doi.org/10.1088/1367-2630/14/3/035019} {\bibfield  {journal}
  {\bibinfo  {journal} {New J. Phys.}\ }\textbf {\bibinfo {volume} {14}},\
  \bibinfo {pages} {035019} (\bibinfo {year} {2012})}\BibitemShut {NoStop}%
\bibitem [{\citenamefont {Hyart}\ \emph {et~al.}(2013)\citenamefont {Hyart},
  \citenamefont {van Heck}, \citenamefont {Fulga}, \citenamefont {Burrello},
  \citenamefont {Akhmerov},\ and\ \citenamefont {Beenakker}}]{hyart2013}%
  \BibitemOpen
  \bibfield  {author} {\bibinfo {author} {\bibfnamefont {T.}~\bibnamefont
  {Hyart}}, \bibinfo {author} {\bibfnamefont {B.}~\bibnamefont {van Heck}},
  \bibinfo {author} {\bibfnamefont {I.~C.}\ \bibnamefont {Fulga}}, \bibinfo
  {author} {\bibfnamefont {M.}~\bibnamefont {Burrello}}, \bibinfo {author}
  {\bibfnamefont {A.~R.}\ \bibnamefont {Akhmerov}},\ and\ \bibinfo {author}
  {\bibfnamefont {C.~W.~J.}\ \bibnamefont {Beenakker}},\ }\bibfield  {title}
  {\bibinfo {title} {{Flux-controlled quantum computation with Majorana
  fermions}},\ }\href {https://doi.org/10.1103/PhysRevB.88.035121} {\bibfield
  {journal} {\bibinfo  {journal} {Phys. Rev. B}\ }\textbf {\bibinfo {volume}
  {88}},\ \bibinfo {pages} {035121} (\bibinfo {year} {2013})}\BibitemShut
  {NoStop}%
\bibitem [{\citenamefont {Das~Sarma}\ \emph {et~al.}(2015)\citenamefont
  {Das~Sarma}, \citenamefont {Freedman},\ and\ \citenamefont
  {Nayak}}]{sarma2015majorana}%
  \BibitemOpen
  \bibfield  {author} {\bibinfo {author} {\bibfnamefont {S.}~\bibnamefont
  {Das~Sarma}}, \bibinfo {author} {\bibfnamefont {M.}~\bibnamefont
  {Freedman}},\ and\ \bibinfo {author} {\bibfnamefont {C.}~\bibnamefont
  {Nayak}},\ }\bibfield  {title} {\bibinfo {title} {{Majorana zero modes and
  topological quantum computation}},\ }\href
  {https://doi.org/10.1038/npjqi.2015.1} {\bibfield  {journal} {\bibinfo
  {journal} {npj Quantum Inf.}\ }\textbf {\bibinfo {volume} {1}},\ \bibinfo
  {pages} {15001} (\bibinfo {year} {2015})}\BibitemShut {NoStop}%
\bibitem [{\citenamefont {Aasen}\ \emph {et~al.}(2016)\citenamefont {Aasen},
  \citenamefont {Hell}, \citenamefont {Mishmash}, \citenamefont {Higginbotham},
  \citenamefont {Danon}, \citenamefont {Leijnse}, \citenamefont {Jespersen},
  \citenamefont {Folk}, \citenamefont {Marcus}, \citenamefont {Flensberg},\
  and\ \citenamefont {Alicea}}]{aasen2016milestones}%
  \BibitemOpen
  \bibfield  {author} {\bibinfo {author} {\bibfnamefont {D.}~\bibnamefont
  {Aasen}}, \bibinfo {author} {\bibfnamefont {M.}~\bibnamefont {Hell}},
  \bibinfo {author} {\bibfnamefont {R.~V.}\ \bibnamefont {Mishmash}}, \bibinfo
  {author} {\bibfnamefont {A.}~\bibnamefont {Higginbotham}}, \bibinfo {author}
  {\bibfnamefont {J.}~\bibnamefont {Danon}}, \bibinfo {author} {\bibfnamefont
  {M.}~\bibnamefont {Leijnse}}, \bibinfo {author} {\bibfnamefont {T.~S.}\
  \bibnamefont {Jespersen}}, \bibinfo {author} {\bibfnamefont {J.~A.}\
  \bibnamefont {Folk}}, \bibinfo {author} {\bibfnamefont {C.~M.}\ \bibnamefont
  {Marcus}}, \bibinfo {author} {\bibfnamefont {K.}~\bibnamefont {Flensberg}},\
  and\ \bibinfo {author} {\bibfnamefont {J.}~\bibnamefont {Alicea}},\
  }\bibfield  {title} {\bibinfo {title} {{Milestones Toward Majorana-Based
  Quantum Computing}},\ }\href {https://doi.org/10.1103/PhysRevX.6.031016}
  {\bibfield  {journal} {\bibinfo  {journal} {Phys. Rev. X}\ }\textbf {\bibinfo
  {volume} {6}},\ \bibinfo {pages} {031016} (\bibinfo {year}
  {2016})}\BibitemShut {NoStop}%
\bibitem [{\citenamefont {Plugge}\ \emph {et~al.}(2017)\citenamefont {Plugge},
  \citenamefont {Rasmussen}, \citenamefont {Egger},\ and\ \citenamefont
  {Flensberg}}]{plugge2017majorana}%
  \BibitemOpen
  \bibfield  {author} {\bibinfo {author} {\bibfnamefont {S.}~\bibnamefont
  {Plugge}}, \bibinfo {author} {\bibfnamefont {A.}~\bibnamefont {Rasmussen}},
  \bibinfo {author} {\bibfnamefont {R.}~\bibnamefont {Egger}},\ and\ \bibinfo
  {author} {\bibfnamefont {K.}~\bibnamefont {Flensberg}},\ }\bibfield  {title}
  {\bibinfo {title} {{Majorana box qubits}},\ }\href
  {https://doi.org/10.1088/1367-2630/aa54e1} {\bibfield  {journal} {\bibinfo
  {journal} {New J. Phys.}\ }\textbf {\bibinfo {volume} {19}},\ \bibinfo
  {pages} {012001} (\bibinfo {year} {2017})}\BibitemShut {NoStop}%
\bibitem [{\citenamefont {Karzig}\ \emph {et~al.}(2017)\citenamefont {Karzig},
  \citenamefont {Knapp}, \citenamefont {Lutchyn}, \citenamefont {Bonderson},
  \citenamefont {Hastings}, \citenamefont {Nayak}, \citenamefont {Alicea},
  \citenamefont {Flensberg}, \citenamefont {Plugge}, \citenamefont {Oreg},
  \citenamefont {Marcus},\ and\ \citenamefont {Freedman}}]{karzig2017scalable}%
  \BibitemOpen
  \bibfield  {author} {\bibinfo {author} {\bibfnamefont {T.}~\bibnamefont
  {Karzig}}, \bibinfo {author} {\bibfnamefont {C.}~\bibnamefont {Knapp}},
  \bibinfo {author} {\bibfnamefont {R.~M.}\ \bibnamefont {Lutchyn}}, \bibinfo
  {author} {\bibfnamefont {P.}~\bibnamefont {Bonderson}}, \bibinfo {author}
  {\bibfnamefont {M.~B.}\ \bibnamefont {Hastings}}, \bibinfo {author}
  {\bibfnamefont {C.}~\bibnamefont {Nayak}}, \bibinfo {author} {\bibfnamefont
  {J.}~\bibnamefont {Alicea}}, \bibinfo {author} {\bibfnamefont
  {K.}~\bibnamefont {Flensberg}}, \bibinfo {author} {\bibfnamefont
  {S.}~\bibnamefont {Plugge}}, \bibinfo {author} {\bibfnamefont
  {Y.}~\bibnamefont {Oreg}}, \bibinfo {author} {\bibfnamefont {C.~M.}\
  \bibnamefont {Marcus}},\ and\ \bibinfo {author} {\bibfnamefont {M.~H.}\
  \bibnamefont {Freedman}},\ }\bibfield  {title} {\bibinfo {title} {{Scalable
  designs for quasiparticle-poisoning-protected topological quantum computation
  with Majorana zero modes}},\ }\href
  {https://doi.org/10.1103/PhysRevB.95.235305} {\bibfield  {journal} {\bibinfo
  {journal} {Phys. Rev. B}\ }\textbf {\bibinfo {volume} {95}},\ \bibinfo
  {pages} {235305} (\bibinfo {year} {2017})}\BibitemShut {NoStop}%
\bibitem [{\citenamefont {Ohm}\ and\ \citenamefont
  {Hassler}(2015)}]{ohm2015microwave}%
  \BibitemOpen
  \bibfield  {author} {\bibinfo {author} {\bibfnamefont {C.}~\bibnamefont
  {Ohm}}\ and\ \bibinfo {author} {\bibfnamefont {F.}~\bibnamefont {Hassler}},\
  }\bibfield  {title} {\bibinfo {title} {{Microwave readout of Majorana
  qubits}},\ }\href {https://doi.org/10.1103/PhysRevB.91.085406} {\bibfield
  {journal} {\bibinfo  {journal} {Phys. Rev. B}\ }\textbf {\bibinfo {volume}
  {91}},\ \bibinfo {pages} {085406} (\bibinfo {year} {2015})}\BibitemShut
  {NoStop}%
\bibitem [{\citenamefont {Gharavi}\ \emph {et~al.}(2016)\citenamefont
  {Gharavi}, \citenamefont {Hoving},\ and\ \citenamefont
  {Baugh}}]{gharavi2016readout}%
  \BibitemOpen
  \bibfield  {author} {\bibinfo {author} {\bibfnamefont {K.}~\bibnamefont
  {Gharavi}}, \bibinfo {author} {\bibfnamefont {D.}~\bibnamefont {Hoving}},\
  and\ \bibinfo {author} {\bibfnamefont {J.}~\bibnamefont {Baugh}},\ }\bibfield
   {title} {\bibinfo {title} {{Readout of Majorana parity states using a
  quantum dot}},\ }\href {https://doi.org/10.1103/PhysRevB.94.155417}
  {\bibfield  {journal} {\bibinfo  {journal} {Phys. Rev. B}\ }\textbf {\bibinfo
  {volume} {94}},\ \bibinfo {pages} {155417} (\bibinfo {year}
  {2016})}\BibitemShut {NoStop}%
\bibitem [{\citenamefont {Li}\ \emph {et~al.}(2018)\citenamefont {Li},
  \citenamefont {Coish}, \citenamefont {Hell}, \citenamefont {Flensberg},\ and\
  \citenamefont {Leijnse}}]{li2018four}%
  \BibitemOpen
  \bibfield  {author} {\bibinfo {author} {\bibfnamefont {T.}~\bibnamefont
  {Li}}, \bibinfo {author} {\bibfnamefont {W.~A.}\ \bibnamefont {Coish}},
  \bibinfo {author} {\bibfnamefont {M.}~\bibnamefont {Hell}}, \bibinfo {author}
  {\bibfnamefont {K.}~\bibnamefont {Flensberg}},\ and\ \bibinfo {author}
  {\bibfnamefont {M.}~\bibnamefont {Leijnse}},\ }\bibfield  {title} {\bibinfo
  {title} {{Four-Majorana qubit with charge readout: Dynamics and
  decoherence}},\ }\href {https://doi.org/10.1103/PhysRevB.98.205403}
  {\bibfield  {journal} {\bibinfo  {journal} {Phys. Rev. B}\ }\textbf {\bibinfo
  {volume} {98}},\ \bibinfo {pages} {205403} (\bibinfo {year}
  {2018})}\BibitemShut {NoStop}%
\bibitem [{\citenamefont {Grimsmo}\ and\ \citenamefont
  {Smith}(2019)}]{grimsmo2019majorana}%
  \BibitemOpen
  \bibfield  {author} {\bibinfo {author} {\bibfnamefont {A.~L.}\ \bibnamefont
  {Grimsmo}}\ and\ \bibinfo {author} {\bibfnamefont {T.~B.}\ \bibnamefont
  {Smith}},\ }\bibfield  {title} {\bibinfo {title} {{Majorana qubit readout
  using longitudinal qubit-resonator interaction}},\ }\href
  {https://doi.org/10.1103/PhysRevB.99.235420} {\bibfield  {journal} {\bibinfo
  {journal} {Phys. Rev. B}\ }\textbf {\bibinfo {volume} {99}},\ \bibinfo
  {pages} {235420} (\bibinfo {year} {2019})}\BibitemShut {NoStop}%
\bibitem [{\citenamefont {Sz\'echenyi}\ and\ \citenamefont
  {P\'alyi}(2020)}]{Szechenyi2020}%
  \BibitemOpen
  \bibfield  {author} {\bibinfo {author} {\bibfnamefont {G.}~\bibnamefont
  {Sz\'echenyi}}\ and\ \bibinfo {author} {\bibfnamefont {A.}~\bibnamefont
  {P\'alyi}},\ }\bibfield  {title} {\bibinfo {title} {Parity-to-charge
  conversion for readout of topological majorana qubits},\ }\href
  {https://doi.org/10.1103/PhysRevB.101.235441} {\bibfield  {journal} {\bibinfo
   {journal} {Phys. Rev. B}\ }\textbf {\bibinfo {volume} {101}},\ \bibinfo
  {pages} {235441} (\bibinfo {year} {2020})}\BibitemShut {NoStop}%
\bibitem [{\citenamefont {Steiner}\ and\ \citenamefont {von
  Oppen}(2020)}]{steiner2020readout}%
  \BibitemOpen
  \bibfield  {author} {\bibinfo {author} {\bibfnamefont {J.~F.}\ \bibnamefont
  {Steiner}}\ and\ \bibinfo {author} {\bibfnamefont {F.}~\bibnamefont {von
  Oppen}},\ }\bibfield  {title} {\bibinfo {title} {Readout of majorana
  qubits},\ }\href {https://doi.org/10.1103/PhysRevResearch.2.033255}
  {\bibfield  {journal} {\bibinfo  {journal} {Phys. Rev. Research}\ }\textbf
  {\bibinfo {volume} {2}},\ \bibinfo {pages} {033255} (\bibinfo {year}
  {2020})}\BibitemShut {NoStop}%
\bibitem [{\citenamefont {Munk}\ \emph {et~al.}(2020)\citenamefont {Munk},
  \citenamefont {Schulenborg}, \citenamefont {Egger},\ and\ \citenamefont
  {Flensberg}}]{munk2020paritytocharge}%
  \BibitemOpen
  \bibfield  {author} {\bibinfo {author} {\bibfnamefont {M.~I.~K.}\
  \bibnamefont {Munk}}, \bibinfo {author} {\bibfnamefont {J.}~\bibnamefont
  {Schulenborg}}, \bibinfo {author} {\bibfnamefont {R.}~\bibnamefont {Egger}},\
  and\ \bibinfo {author} {\bibfnamefont {K.}~\bibnamefont {Flensberg}},\
  }\bibfield  {title} {\bibinfo {title} {Parity-to-charge conversion in
  majorana qubit readout},\ }\href
  {https://doi.org/10.1103/PhysRevResearch.2.033254} {\bibfield  {journal}
  {\bibinfo  {journal} {Phys. Rev. Research}\ }\textbf {\bibinfo {volume}
  {2}},\ \bibinfo {pages} {033254} (\bibinfo {year} {2020})}\BibitemShut
  {NoStop}%
\bibitem [{\citenamefont {Maman}\ \emph {et~al.}(2020)\citenamefont {Maman},
  \citenamefont {Gonzalez-Zalba},\ and\ \citenamefont
  {P{\'a}lyi}}]{maman2020charge}%
  \BibitemOpen
  \bibfield  {author} {\bibinfo {author} {\bibfnamefont {V.~D.}\ \bibnamefont
  {Maman}}, \bibinfo {author} {\bibfnamefont {M.}~\bibnamefont
  {Gonzalez-Zalba}},\ and\ \bibinfo {author} {\bibfnamefont {A.}~\bibnamefont
  {P{\'a}lyi}},\ }\bibfield  {title} {\bibinfo {title} {Charge noise and
  overdrive errors in reflectometry-based charge, spin and majorana qubit
  readout},\ }\Eprint {https://arxiv.org/abs/2006.12391} {arXiv:2006.12391
  [cond-mat.mes-hall]}  (\bibinfo {year} {2020})\BibitemShut {NoStop}%
\bibitem [{\citenamefont {Khindanov}\ \emph {et~al.}(2020)\citenamefont
  {Khindanov}, \citenamefont {Pikulin},\ and\ \citenamefont
  {Karzig}}]{khindanov2020visibility}%
  \BibitemOpen
  \bibfield  {author} {\bibinfo {author} {\bibfnamefont {A.}~\bibnamefont
  {Khindanov}}, \bibinfo {author} {\bibfnamefont {D.}~\bibnamefont {Pikulin}},\
  and\ \bibinfo {author} {\bibfnamefont {T.}~\bibnamefont {Karzig}},\
  }\bibfield  {title} {\bibinfo {title} {{Visibility of noisy quantum dot-based
  measurements of Majorana qubits}},\ }\Eprint
  {https://arxiv.org/abs/2007.11024} {arXiv:2007.11024 [cond-mat.mes-hall]}
  (\bibinfo {year} {2020})\BibitemShut {NoStop}%
\bibitem [{\citenamefont {Bonderson}\ \emph {et~al.}(2008)\citenamefont
  {Bonderson}, \citenamefont {Freedman},\ and\ \citenamefont
  {Nayak}}]{bonderson2008measurement}%
  \BibitemOpen
  \bibfield  {author} {\bibinfo {author} {\bibfnamefont {P.}~\bibnamefont
  {Bonderson}}, \bibinfo {author} {\bibfnamefont {M.}~\bibnamefont
  {Freedman}},\ and\ \bibinfo {author} {\bibfnamefont {C.}~\bibnamefont
  {Nayak}},\ }\bibfield  {title} {\bibinfo {title} {{Measurement-Only
  Topological Quantum Computation}},\ }\href
  {https://doi.org/10.1103/PhysRevLett.101.010501} {\bibfield  {journal}
  {\bibinfo  {journal} {Phys. Rev. Lett.}\ }\textbf {\bibinfo {volume} {101}},\
  \bibinfo {pages} {010501} (\bibinfo {year} {2008})}\BibitemShut {NoStop}%
\bibitem [{\citenamefont {Bonderson}\ \emph {et~al.}(2009)\citenamefont
  {Bonderson}, \citenamefont {Freedman},\ and\ \citenamefont
  {Nayak}}]{bonderson2009measurement}%
  \BibitemOpen
  \bibfield  {author} {\bibinfo {author} {\bibfnamefont {P.}~\bibnamefont
  {Bonderson}}, \bibinfo {author} {\bibfnamefont {M.}~\bibnamefont
  {Freedman}},\ and\ \bibinfo {author} {\bibfnamefont {C.}~\bibnamefont
  {Nayak}},\ }\bibfield  {title} {\bibinfo {title} {{Measurement-only
  topological quantum computation via anyonic interferometry}},\ }\href
  {https://doi.org/10.1016/j.aop.2008.09.009} {\bibfield  {journal} {\bibinfo
  {journal} {Ann. Phys.}\ }\textbf {\bibinfo {volume} {324}},\ \bibinfo {pages}
  {787–826} (\bibinfo {year} {2009})}\BibitemShut {NoStop}%
\bibitem [{\citenamefont {Karzig}\ \emph {et~al.}(2019)\citenamefont {Karzig},
  \citenamefont {Oreg}, \citenamefont {Refael},\ and\ \citenamefont
  {Freedman}}]{karzig2019robust}%
  \BibitemOpen
  \bibfield  {author} {\bibinfo {author} {\bibfnamefont {T.}~\bibnamefont
  {Karzig}}, \bibinfo {author} {\bibfnamefont {Y.}~\bibnamefont {Oreg}},
  \bibinfo {author} {\bibfnamefont {G.}~\bibnamefont {Refael}},\ and\ \bibinfo
  {author} {\bibfnamefont {M.~H.}\ \bibnamefont {Freedman}},\ }\bibfield
  {title} {\bibinfo {title} {Robust majorana magic gates via measurements},\
  }\href {https://doi.org/10.1103/PhysRevB.99.144521} {\bibfield  {journal}
  {\bibinfo  {journal} {Phys. Rev. B}\ }\textbf {\bibinfo {volume} {99}},\
  \bibinfo {pages} {144521} (\bibinfo {year} {2019})}\BibitemShut {NoStop}%
\bibitem [{\citenamefont {Blais}\ \emph {et~al.}(2020)\citenamefont {Blais},
  \citenamefont {Grimsmo}, \citenamefont {Girvin},\ and\ \citenamefont
  {Wallraff}}]{blais2020circuit}%
  \BibitemOpen
  \bibfield  {author} {\bibinfo {author} {\bibfnamefont {A.}~\bibnamefont
  {Blais}}, \bibinfo {author} {\bibfnamefont {A.~L.}\ \bibnamefont {Grimsmo}},
  \bibinfo {author} {\bibfnamefont {S.~M.}\ \bibnamefont {Girvin}},\ and\
  \bibinfo {author} {\bibfnamefont {A.}~\bibnamefont {Wallraff}},\ }\bibfield
  {title} {\bibinfo {title} {{Circuit Quantum Electrodynamics}},\ }\Eprint
  {https://arxiv.org/abs/2005.12667} {arXiv:2005.12667 [quant-ph]}  (\bibinfo
  {year} {2020})\BibitemShut {NoStop}%
\bibitem [{\citenamefont {Wallraff}\ \emph {et~al.}(2004)\citenamefont
  {Wallraff}, \citenamefont {Schuster}, \citenamefont {Blais}, \citenamefont
  {Frunzio}, \citenamefont {Huang}, \citenamefont {Majer}, \citenamefont
  {Kumar}, \citenamefont {Girvin},\ and\ \citenamefont
  {Schoelkopf}}]{wallraff2004circuit}%
  \BibitemOpen
  \bibfield  {author} {\bibinfo {author} {\bibfnamefont {A.}~\bibnamefont
  {Wallraff}}, \bibinfo {author} {\bibfnamefont {D.~I.}\ \bibnamefont
  {Schuster}}, \bibinfo {author} {\bibfnamefont {A.}~\bibnamefont {Blais}},
  \bibinfo {author} {\bibfnamefont {L.}~\bibnamefont {Frunzio}}, \bibinfo
  {author} {\bibfnamefont {R.-S.}\ \bibnamefont {Huang}}, \bibinfo {author}
  {\bibfnamefont {J.}~\bibnamefont {Majer}}, \bibinfo {author} {\bibfnamefont
  {S.}~\bibnamefont {Kumar}}, \bibinfo {author} {\bibfnamefont {S.~M.}\
  \bibnamefont {Girvin}},\ and\ \bibinfo {author} {\bibfnamefont {R.~J.}\
  \bibnamefont {Schoelkopf}},\ }\bibfield  {title} {\bibinfo {title} {{Strong
  coupling of a single photon to a superconducting qubit using circuit quantum
  electrodynamics}},\ }\href {https://doi.org/10.1038/nature02851} {\bibfield
  {journal} {\bibinfo  {journal} {Nature}\ }\textbf {\bibinfo {volume} {431}},\
  \bibinfo {pages} {162–167} (\bibinfo {year} {2004})}\BibitemShut {NoStop}%
\bibitem [{\citenamefont {Walter}\ \emph {et~al.}(2017)\citenamefont {Walter},
  \citenamefont {Kurpiers}, \citenamefont {Gasparinetti}, \citenamefont
  {Magnard}, \citenamefont {Poto\ifmmode~\check{c}\else \v{c}\fi{}nik},
  \citenamefont {Salath\'e}, \citenamefont {Pechal}, \citenamefont {Mondal},
  \citenamefont {Oppliger}, \citenamefont {Eichler},\ and\ \citenamefont
  {Wallraff}}]{walter2017rapid}%
  \BibitemOpen
  \bibfield  {author} {\bibinfo {author} {\bibfnamefont {T.}~\bibnamefont
  {Walter}}, \bibinfo {author} {\bibfnamefont {P.}~\bibnamefont {Kurpiers}},
  \bibinfo {author} {\bibfnamefont {S.}~\bibnamefont {Gasparinetti}}, \bibinfo
  {author} {\bibfnamefont {P.}~\bibnamefont {Magnard}}, \bibinfo {author}
  {\bibfnamefont {A.}~\bibnamefont {Poto\ifmmode~\check{c}\else
  \v{c}\fi{}nik}}, \bibinfo {author} {\bibfnamefont {Y.}~\bibnamefont
  {Salath\'e}}, \bibinfo {author} {\bibfnamefont {M.}~\bibnamefont {Pechal}},
  \bibinfo {author} {\bibfnamefont {M.}~\bibnamefont {Mondal}}, \bibinfo
  {author} {\bibfnamefont {M.}~\bibnamefont {Oppliger}}, \bibinfo {author}
  {\bibfnamefont {C.}~\bibnamefont {Eichler}},\ and\ \bibinfo {author}
  {\bibfnamefont {A.}~\bibnamefont {Wallraff}},\ }\bibfield  {title} {\bibinfo
  {title} {{Rapid High-Fidelity Single-Shot Dispersive Readout of
  Superconducting Qubits}},\ }\href
  {https://doi.org/10.1103/PhysRevApplied.7.054020} {\bibfield  {journal}
  {\bibinfo  {journal} {Phys. Rev. Applied}\ }\textbf {\bibinfo {volume} {7}},\
  \bibinfo {pages} {054020} (\bibinfo {year} {2017})}\BibitemShut {NoStop}%
\bibitem [{\citenamefont {Colless}\ \emph {et~al.}(2013)\citenamefont
  {Colless}, \citenamefont {Mahoney}, \citenamefont {Hornibrook}, \citenamefont
  {Doherty}, \citenamefont {Lu}, \citenamefont {Gossard},\ and\ \citenamefont
  {Reilly}}]{colless2013dispersive}%
  \BibitemOpen
  \bibfield  {author} {\bibinfo {author} {\bibfnamefont {J.~I.}\ \bibnamefont
  {Colless}}, \bibinfo {author} {\bibfnamefont {A.~C.}\ \bibnamefont
  {Mahoney}}, \bibinfo {author} {\bibfnamefont {J.~M.}\ \bibnamefont
  {Hornibrook}}, \bibinfo {author} {\bibfnamefont {A.~C.}\ \bibnamefont
  {Doherty}}, \bibinfo {author} {\bibfnamefont {H.}~\bibnamefont {Lu}},
  \bibinfo {author} {\bibfnamefont {A.~C.}\ \bibnamefont {Gossard}},\ and\
  \bibinfo {author} {\bibfnamefont {D.~J.}\ \bibnamefont {Reilly}},\ }\bibfield
   {title} {\bibinfo {title} {{Dispersive Readout of a Few-Electron Double
  Quantum Dot with Fast rf Gate Sensors}},\ }\href
  {https://doi.org/10.1103/PhysRevLett.110.046805} {\bibfield  {journal}
  {\bibinfo  {journal} {Phys. Rev. Lett.}\ }\textbf {\bibinfo {volume} {110}},\
  \bibinfo {pages} {046805} (\bibinfo {year} {2013})}\BibitemShut {NoStop}%
\bibitem [{\citenamefont {West}\ \emph {et~al.}(2019)\citenamefont {West},
  \citenamefont {Hensen}, \citenamefont {Jouan}, \citenamefont {Tanttu},
  \citenamefont {Yang}, \citenamefont {Rossi}, \citenamefont {Gonzalez-Zalba},
  \citenamefont {Hudson}, \citenamefont {Morello}, \citenamefont {Reilly},\
  and\ \citenamefont {Dzurak}}]{west2019gate}%
  \BibitemOpen
  \bibfield  {author} {\bibinfo {author} {\bibfnamefont {A.}~\bibnamefont
  {West}}, \bibinfo {author} {\bibfnamefont {B.}~\bibnamefont {Hensen}},
  \bibinfo {author} {\bibfnamefont {A.}~\bibnamefont {Jouan}}, \bibinfo
  {author} {\bibfnamefont {T.}~\bibnamefont {Tanttu}}, \bibinfo {author}
  {\bibfnamefont {C.-H.}\ \bibnamefont {Yang}}, \bibinfo {author}
  {\bibfnamefont {A.}~\bibnamefont {Rossi}}, \bibinfo {author} {\bibfnamefont
  {M.~F.}\ \bibnamefont {Gonzalez-Zalba}}, \bibinfo {author} {\bibfnamefont
  {F.}~\bibnamefont {Hudson}}, \bibinfo {author} {\bibfnamefont
  {A.}~\bibnamefont {Morello}}, \bibinfo {author} {\bibfnamefont {D.~J.}\
  \bibnamefont {Reilly}},\ and\ \bibinfo {author} {\bibfnamefont {A.~S.}\
  \bibnamefont {Dzurak}},\ }\bibfield  {title} {\bibinfo {title} {{Gate-based
  single-shot readout of spins in silicon}},\ }\href
  {https://doi.org/10.1038/s41565-019-0400-7} {\bibfield  {journal} {\bibinfo
  {journal} {Nat. Nanotechnol.}\ }\textbf {\bibinfo {volume} {14}},\ \bibinfo
  {pages} {437–441} (\bibinfo {year} {2019})}\BibitemShut {NoStop}%
\bibitem [{\citenamefont {de~Jong}\ \emph {et~al.}(2019)\citenamefont
  {de~Jong}, \citenamefont {van Veen}, \citenamefont {Binci}, \citenamefont
  {Singh}, \citenamefont {Krogstrup}, \citenamefont {Kouwenhoven},
  \citenamefont {Pfaff},\ and\ \citenamefont {Watson}}]{de2019rapid}%
  \BibitemOpen
  \bibfield  {author} {\bibinfo {author} {\bibfnamefont {D.}~\bibnamefont
  {de~Jong}}, \bibinfo {author} {\bibfnamefont {J.}~\bibnamefont {van Veen}},
  \bibinfo {author} {\bibfnamefont {L.}~\bibnamefont {Binci}}, \bibinfo
  {author} {\bibfnamefont {A.}~\bibnamefont {Singh}}, \bibinfo {author}
  {\bibfnamefont {P.}~\bibnamefont {Krogstrup}}, \bibinfo {author}
  {\bibfnamefont {L.~P.}\ \bibnamefont {Kouwenhoven}}, \bibinfo {author}
  {\bibfnamefont {W.}~\bibnamefont {Pfaff}},\ and\ \bibinfo {author}
  {\bibfnamefont {J.~D.}\ \bibnamefont {Watson}},\ }\bibfield  {title}
  {\bibinfo {title} {{Rapid Detection of Coherent Tunneling in an {InAs}
  Nanowire Quantum Dot through Dispersive Gate Sensing}},\ }\href
  {https://doi.org/10.1103/PhysRevApplied.11.044061} {\bibfield  {journal}
  {\bibinfo  {journal} {Phys. Rev. Applied}\ }\textbf {\bibinfo {volume}
  {11}},\ \bibinfo {pages} {044061} (\bibinfo {year} {2019})}\BibitemShut
  {NoStop}%
\bibitem [{\citenamefont {Larsen}\ \emph {et~al.}(2015)\citenamefont {Larsen},
  \citenamefont {Petersson}, \citenamefont {Kuemmeth}, \citenamefont
  {Jespersen}, \citenamefont {Krogstrup}, \citenamefont {Nyg\aa{}rd},\ and\
  \citenamefont {Marcus}}]{larsen2015semiconductor}%
  \BibitemOpen
  \bibfield  {author} {\bibinfo {author} {\bibfnamefont {T.~W.}\ \bibnamefont
  {Larsen}}, \bibinfo {author} {\bibfnamefont {K.~D.}\ \bibnamefont
  {Petersson}}, \bibinfo {author} {\bibfnamefont {F.}~\bibnamefont {Kuemmeth}},
  \bibinfo {author} {\bibfnamefont {T.~S.}\ \bibnamefont {Jespersen}}, \bibinfo
  {author} {\bibfnamefont {P.}~\bibnamefont {Krogstrup}}, \bibinfo {author}
  {\bibfnamefont {J.}~\bibnamefont {Nyg\aa{}rd}},\ and\ \bibinfo {author}
  {\bibfnamefont {C.~M.}\ \bibnamefont {Marcus}},\ }\bibfield  {title}
  {\bibinfo {title} {{Semiconductor-Nanowire-Based Superconducting Qubit}},\
  }\href {https://doi.org/10.1103/PhysRevLett.115.127001} {\bibfield  {journal}
  {\bibinfo  {journal} {Phys. Rev. Lett.}\ }\textbf {\bibinfo {volume} {115}},\
  \bibinfo {pages} {127001} (\bibinfo {year} {2015})}\BibitemShut {NoStop}%
\bibitem [{\citenamefont {de~Lange}\ \emph {et~al.}(2015)\citenamefont
  {de~Lange}, \citenamefont {van Heck}, \citenamefont {Bruno}, \citenamefont
  {van Woerkom}, \citenamefont {Geresdi}, \citenamefont {Plissard},
  \citenamefont {Bakkers}, \citenamefont {Akhmerov},\ and\ \citenamefont
  {DiCarlo}}]{de2015realization}%
  \BibitemOpen
  \bibfield  {author} {\bibinfo {author} {\bibfnamefont {G.}~\bibnamefont
  {de~Lange}}, \bibinfo {author} {\bibfnamefont {B.}~\bibnamefont {van Heck}},
  \bibinfo {author} {\bibfnamefont {A.}~\bibnamefont {Bruno}}, \bibinfo
  {author} {\bibfnamefont {D.~J.}\ \bibnamefont {van Woerkom}}, \bibinfo
  {author} {\bibfnamefont {A.}~\bibnamefont {Geresdi}}, \bibinfo {author}
  {\bibfnamefont {S.~R.}\ \bibnamefont {Plissard}}, \bibinfo {author}
  {\bibfnamefont {E.~P. A.~M.}\ \bibnamefont {Bakkers}}, \bibinfo {author}
  {\bibfnamefont {A.~R.}\ \bibnamefont {Akhmerov}},\ and\ \bibinfo {author}
  {\bibfnamefont {L.}~\bibnamefont {DiCarlo}},\ }\bibfield  {title} {\bibinfo
  {title} {{Realization of Microwave Quantum Circuits Using Hybrid
  Superconducting-Semiconducting Nanowire Josephson Elements}},\ }\href
  {https://doi.org/10.1103/PhysRevLett.115.127002} {\bibfield  {journal}
  {\bibinfo  {journal} {Phys. Rev. Lett.}\ }\textbf {\bibinfo {volume} {115}},\
  \bibinfo {pages} {127002} (\bibinfo {year} {2015})}\BibitemShut {NoStop}%
\bibitem [{\citenamefont {Ginossar}\ and\ \citenamefont
  {Grosfeld}(2014)}]{ginossar2014microwave}%
  \BibitemOpen
  \bibfield  {author} {\bibinfo {author} {\bibfnamefont {E.}~\bibnamefont
  {Ginossar}}\ and\ \bibinfo {author} {\bibfnamefont {E.}~\bibnamefont
  {Grosfeld}},\ }\bibfield  {title} {\bibinfo {title} {{Microwave transitions
  as a signature of coherent parity mixing effects in the Majorana-transmon
  qubit}},\ }\href {https://doi.org/10.1038/ncomms5772} {\bibfield  {journal}
  {\bibinfo  {journal} {Nat. Commun.}\ }\textbf {\bibinfo {volume} {5}},\
  \bibinfo {pages} {1} (\bibinfo {year} {2014})}\BibitemShut {NoStop}%
\bibitem [{\citenamefont {Yavilberg}\ \emph {et~al.}(2015)\citenamefont
  {Yavilberg}, \citenamefont {Ginossar},\ and\ \citenamefont
  {Grosfeld}}]{yavilberg2015fermion}%
  \BibitemOpen
  \bibfield  {author} {\bibinfo {author} {\bibfnamefont {K.}~\bibnamefont
  {Yavilberg}}, \bibinfo {author} {\bibfnamefont {E.}~\bibnamefont
  {Ginossar}},\ and\ \bibinfo {author} {\bibfnamefont {E.}~\bibnamefont
  {Grosfeld}},\ }\bibfield  {title} {\bibinfo {title} {{Fermion parity
  measurement and control in Majorana circuit quantum electrodynamics}},\
  }\href {https://doi.org/10.1103/PhysRevB.92.075143} {\bibfield  {journal}
  {\bibinfo  {journal} {Phys. Rev. B}\ }\textbf {\bibinfo {volume} {92}},\
  \bibinfo {pages} {075143} (\bibinfo {year} {2015})}\BibitemShut {NoStop}%
\bibitem [{\citenamefont {Hell}\ \emph {et~al.}(2016)\citenamefont {Hell},
  \citenamefont {Danon}, \citenamefont {Flensberg},\ and\ \citenamefont
  {Leijnse}}]{hell2016time}%
  \BibitemOpen
  \bibfield  {author} {\bibinfo {author} {\bibfnamefont {M.}~\bibnamefont
  {Hell}}, \bibinfo {author} {\bibfnamefont {J.}~\bibnamefont {Danon}},
  \bibinfo {author} {\bibfnamefont {K.}~\bibnamefont {Flensberg}},\ and\
  \bibinfo {author} {\bibfnamefont {M.}~\bibnamefont {Leijnse}},\ }\bibfield
  {title} {\bibinfo {title} {{Time scales for Majorana manipulation using
  Coulomb blockade in gate-controlled superconducting nanowires}},\ }\href
  {https://doi.org/10.1103/PhysRevB.94.035424} {\bibfield  {journal} {\bibinfo
  {journal} {Phys. Rev. B}\ }\textbf {\bibinfo {volume} {94}},\ \bibinfo
  {pages} {035424} (\bibinfo {year} {2016})}\BibitemShut {NoStop}%
\bibitem [{\citenamefont {Blais}\ \emph {et~al.}(2004)\citenamefont {Blais},
  \citenamefont {Huang}, \citenamefont {Wallraff}, \citenamefont {Girvin},\
  and\ \citenamefont {Schoelkopf}}]{blais2004cavity}%
  \BibitemOpen
  \bibfield  {author} {\bibinfo {author} {\bibfnamefont {A.}~\bibnamefont
  {Blais}}, \bibinfo {author} {\bibfnamefont {R.-S.}\ \bibnamefont {Huang}},
  \bibinfo {author} {\bibfnamefont {A.}~\bibnamefont {Wallraff}}, \bibinfo
  {author} {\bibfnamefont {S.~M.}\ \bibnamefont {Girvin}},\ and\ \bibinfo
  {author} {\bibfnamefont {R.~J.}\ \bibnamefont {Schoelkopf}},\ }\bibfield
  {title} {\bibinfo {title} {{Cavity quantum electrodynamics for
  superconducting electrical circuits: An architecture for quantum
  computation}},\ }\href {https://doi.org/10.1103/PhysRevA.69.062320}
  {\bibfield  {journal} {\bibinfo  {journal} {Phys. Rev. A}\ }\textbf {\bibinfo
  {volume} {69}},\ \bibinfo {pages} {062320} (\bibinfo {year}
  {2004})}\BibitemShut {NoStop}%
\bibitem [{\citenamefont {Koch}\ \emph {et~al.}(2007)\citenamefont {Koch},
  \citenamefont {Yu}, \citenamefont {Gambetta}, \citenamefont {Houck},
  \citenamefont {Schuster}, \citenamefont {Majer}, \citenamefont {Blais},
  \citenamefont {Devoret}, \citenamefont {Girvin},\ and\ \citenamefont
  {Schoelkopf}}]{koch2007charge}%
  \BibitemOpen
  \bibfield  {author} {\bibinfo {author} {\bibfnamefont {J.}~\bibnamefont
  {Koch}}, \bibinfo {author} {\bibfnamefont {T.~M.}\ \bibnamefont {Yu}},
  \bibinfo {author} {\bibfnamefont {J.}~\bibnamefont {Gambetta}}, \bibinfo
  {author} {\bibfnamefont {A.~A.}\ \bibnamefont {Houck}}, \bibinfo {author}
  {\bibfnamefont {D.~I.}\ \bibnamefont {Schuster}}, \bibinfo {author}
  {\bibfnamefont {J.}~\bibnamefont {Majer}}, \bibinfo {author} {\bibfnamefont
  {A.}~\bibnamefont {Blais}}, \bibinfo {author} {\bibfnamefont {M.~H.}\
  \bibnamefont {Devoret}}, \bibinfo {author} {\bibfnamefont {S.~M.}\
  \bibnamefont {Girvin}},\ and\ \bibinfo {author} {\bibfnamefont {R.~J.}\
  \bibnamefont {Schoelkopf}},\ }\bibfield  {title} {\bibinfo {title}
  {{Charge-insensitive qubit design derived from the Cooper pair box}},\ }\href
  {https://doi.org/10.1103/PhysRevA.76.042319} {\bibfield  {journal} {\bibinfo
  {journal} {Phys. Rev. A}\ }\textbf {\bibinfo {volume} {76}},\ \bibinfo
  {pages} {042319} (\bibinfo {year} {2007})}\BibitemShut {NoStop}%
\bibitem [{\citenamefont {Knapp}\ \emph {et~al.}(2018)\citenamefont {Knapp},
  \citenamefont {Karzig}, \citenamefont {Lutchyn},\ and\ \citenamefont
  {Nayak}}]{knapp2018dephasing}%
  \BibitemOpen
  \bibfield  {author} {\bibinfo {author} {\bibfnamefont {C.}~\bibnamefont
  {Knapp}}, \bibinfo {author} {\bibfnamefont {T.}~\bibnamefont {Karzig}},
  \bibinfo {author} {\bibfnamefont {R.~M.}\ \bibnamefont {Lutchyn}},\ and\
  \bibinfo {author} {\bibfnamefont {C.}~\bibnamefont {Nayak}},\ }\bibfield
  {title} {\bibinfo {title} {{Dephasing of Majorana-based qubits}},\ }\href
  {https://doi.org/10.1103/PhysRevB.97.125404} {\bibfield  {journal} {\bibinfo
  {journal} {Phys. Rev. B}\ }\textbf {\bibinfo {volume} {97}},\ \bibinfo
  {pages} {125404} (\bibinfo {year} {2018})}\BibitemShut {NoStop}%
\bibitem [{\citenamefont {Karzig}\ \emph {et~al.}(2020)\citenamefont {Karzig},
  \citenamefont {Cole},\ and\ \citenamefont
  {Pikulin}}]{karzig2020quasiparticle}%
  \BibitemOpen
  \bibfield  {author} {\bibinfo {author} {\bibfnamefont {T.}~\bibnamefont
  {Karzig}}, \bibinfo {author} {\bibfnamefont {W.~S.}\ \bibnamefont {Cole}},\
  and\ \bibinfo {author} {\bibfnamefont {D.~I.}\ \bibnamefont {Pikulin}},\
  }\bibfield  {title} {\bibinfo {title} {{Quasiparticle poisoning of Majorana
  qubits}},\ }\Eprint {https://arxiv.org/abs/2004.01264} {arXiv:2004.01264
  [cond-mat.mes-hall]}  (\bibinfo {year} {2020})\BibitemShut {NoStop}%
\bibitem [{\citenamefont {Deng}\ \emph {et~al.}(2018)\citenamefont {Deng},
  \citenamefont {Vaitiek\ifmmode~\dot{e}\else \.{e}\fi{}nas}, \citenamefont
  {Prada}, \citenamefont {San-Jose}, \citenamefont {Nyg\aa{}rd}, \citenamefont
  {Krogstrup}, \citenamefont {Aguado},\ and\ \citenamefont
  {Marcus}}]{deng2018nonlocality}%
  \BibitemOpen
  \bibfield  {author} {\bibinfo {author} {\bibfnamefont {M.-T.}\ \bibnamefont
  {Deng}}, \bibinfo {author} {\bibfnamefont {S.}~\bibnamefont
  {Vaitiek\ifmmode~\dot{e}\else \.{e}\fi{}nas}}, \bibinfo {author}
  {\bibfnamefont {E.}~\bibnamefont {Prada}}, \bibinfo {author} {\bibfnamefont
  {P.}~\bibnamefont {San-Jose}}, \bibinfo {author} {\bibfnamefont
  {J.}~\bibnamefont {Nyg\aa{}rd}}, \bibinfo {author} {\bibfnamefont
  {P.}~\bibnamefont {Krogstrup}}, \bibinfo {author} {\bibfnamefont
  {R.}~\bibnamefont {Aguado}},\ and\ \bibinfo {author} {\bibfnamefont {C.~M.}\
  \bibnamefont {Marcus}},\ }\bibfield  {title} {\bibinfo {title} {{Nonlocality
  of Majorana modes in hybrid nanowires}},\ }\href
  {https://doi.org/10.1103/PhysRevB.98.085125} {\bibfield  {journal} {\bibinfo
  {journal} {Phys. Rev. B}\ }\textbf {\bibinfo {volume} {98}},\ \bibinfo
  {pages} {085125} (\bibinfo {year} {2018})}\BibitemShut {NoStop}%
\bibitem [{\citenamefont {Houck}\ \emph {et~al.}(2008)\citenamefont {Houck},
  \citenamefont {Schreier}, \citenamefont {Johnson}, \citenamefont {Chow},
  \citenamefont {Koch}, \citenamefont {Gambetta}, \citenamefont {Schuster},
  \citenamefont {Frunzio}, \citenamefont {Devoret}, \citenamefont {Girvin},\
  and\ \citenamefont {Schoelkopf}}]{houck2008controlling}%
  \BibitemOpen
  \bibfield  {author} {\bibinfo {author} {\bibfnamefont {A.~A.}\ \bibnamefont
  {Houck}}, \bibinfo {author} {\bibfnamefont {J.~A.}\ \bibnamefont {Schreier}},
  \bibinfo {author} {\bibfnamefont {B.~R.}\ \bibnamefont {Johnson}}, \bibinfo
  {author} {\bibfnamefont {J.~M.}\ \bibnamefont {Chow}}, \bibinfo {author}
  {\bibfnamefont {J.}~\bibnamefont {Koch}}, \bibinfo {author} {\bibfnamefont
  {J.~M.}\ \bibnamefont {Gambetta}}, \bibinfo {author} {\bibfnamefont {D.~I.}\
  \bibnamefont {Schuster}}, \bibinfo {author} {\bibfnamefont {L.}~\bibnamefont
  {Frunzio}}, \bibinfo {author} {\bibfnamefont {M.~H.}\ \bibnamefont
  {Devoret}}, \bibinfo {author} {\bibfnamefont {S.~M.}\ \bibnamefont
  {Girvin}},\ and\ \bibinfo {author} {\bibfnamefont {R.~J.}\ \bibnamefont
  {Schoelkopf}},\ }\bibfield  {title} {\bibinfo {title} {{Controlling the
  Spontaneous Emission of a Superconducting Transmon Qubit}},\ }\href
  {https://doi.org/10.1103/PhysRevLett.101.080502} {\bibfield  {journal}
  {\bibinfo  {journal} {Phys. Rev. Lett.}\ }\textbf {\bibinfo {volume} {101}},\
  \bibinfo {pages} {080502} (\bibinfo {year} {2008})}\BibitemShut {NoStop}%
\bibitem [{\citenamefont {Fu}(2010)}]{fu2010electron}%
  \BibitemOpen
  \bibfield  {author} {\bibinfo {author} {\bibfnamefont {L.}~\bibnamefont
  {Fu}},\ }\bibfield  {title} {\bibinfo {title} {{Electron Teleportation via
  Majorana Bound States in a Mesoscopic Superconductor}},\ }\href
  {https://doi.org/10.1103/PhysRevLett.104.056402} {\bibfield  {journal}
  {\bibinfo  {journal} {Phys. Rev. Lett.}\ }\textbf {\bibinfo {volume} {104}},\
  \bibinfo {pages} {056402} (\bibinfo {year} {2010})}\BibitemShut {NoStop}%
\bibitem [{\citenamefont {Didier}\ \emph {et~al.}(2015)\citenamefont {Didier},
  \citenamefont {Bourassa},\ and\ \citenamefont {Blais}}]{didier2015fast}%
  \BibitemOpen
  \bibfield  {author} {\bibinfo {author} {\bibfnamefont {N.}~\bibnamefont
  {Didier}}, \bibinfo {author} {\bibfnamefont {J.}~\bibnamefont {Bourassa}},\
  and\ \bibinfo {author} {\bibfnamefont {A.}~\bibnamefont {Blais}},\ }\bibfield
   {title} {\bibinfo {title} {{Fast Quantum Nondemolition Readout by Parametric
  Modulation of Longitudinal Qubit-Oscillator Interaction}},\ }\href
  {https://doi.org/10.1103/PhysRevLett.115.203601} {\bibfield  {journal}
  {\bibinfo  {journal} {Phys. Rev. Lett.}\ }\textbf {\bibinfo {volume} {115}},\
  \bibinfo {pages} {203601} (\bibinfo {year} {2015})}\BibitemShut {NoStop}%
\bibitem [{\citenamefont {Zhu}\ \emph {et~al.}(2013)\citenamefont {Zhu},
  \citenamefont {Ferguson}, \citenamefont {Manucharyan},\ and\ \citenamefont
  {Koch}}]{zhu2013circuit}%
  \BibitemOpen
  \bibfield  {author} {\bibinfo {author} {\bibfnamefont {G.}~\bibnamefont
  {Zhu}}, \bibinfo {author} {\bibfnamefont {D.~G.}\ \bibnamefont {Ferguson}},
  \bibinfo {author} {\bibfnamefont {V.~E.}\ \bibnamefont {Manucharyan}},\ and\
  \bibinfo {author} {\bibfnamefont {J.}~\bibnamefont {Koch}},\ }\bibfield
  {title} {\bibinfo {title} {{Circuit QED with fluxonium qubits: Theory of the
  dispersive regime}},\ }\href {https://doi.org/10.1103/PhysRevB.87.024510}
  {\bibfield  {journal} {\bibinfo  {journal} {Phys. Rev. B}\ }\textbf {\bibinfo
  {volume} {87}},\ \bibinfo {pages} {024510} (\bibinfo {year}
  {2013})}\BibitemShut {NoStop}%
\bibitem [{\citenamefont {Bravyi}\ \emph {et~al.}(2011)\citenamefont {Bravyi},
  \citenamefont {DiVincenzo},\ and\ \citenamefont
  {Loss}}]{bravyi2011schrieffer}%
  \BibitemOpen
  \bibfield  {author} {\bibinfo {author} {\bibfnamefont {S.}~\bibnamefont
  {Bravyi}}, \bibinfo {author} {\bibfnamefont {D.~P.}\ \bibnamefont
  {DiVincenzo}},\ and\ \bibinfo {author} {\bibfnamefont {D.}~\bibnamefont
  {Loss}},\ }\bibfield  {title} {\bibinfo {title} {{Schrieffer–Wolff
  transformation for quantum many-body systems}},\ }\href
  {https://doi.org/10.1016/j.aop.2011.06.004} {\bibfield  {journal} {\bibinfo
  {journal} {Ann. Phys.}\ }\textbf {\bibinfo {volume} {326}},\ \bibinfo {pages}
  {2793–2826} (\bibinfo {year} {2011})}\BibitemShut {NoStop}%
\bibitem [{\citenamefont {Winkler}(2003)}]{winkler2003spin}%
  \BibitemOpen
  \bibfield  {author} {\bibinfo {author} {\bibfnamefont {R.}~\bibnamefont
  {Winkler}},\ }\bibinfo {title} {Quasi-degenerate perturbation theory},\ in\
  \href {https://doi.org/10.1007/978-3-540-36616-4_12} {\emph {\bibinfo
  {booktitle} {Spin--Orbit Coupling Effects in Two-Dimensional Electron and
  Hole Systems}}},\ \bibinfo {series and number} {Springer Tracts in Modern
  Physics}\ (\bibinfo  {publisher} {Springer},\ \bibinfo {year} {2003})\ pp.\
  \bibinfo {pages} {201--206}\BibitemShut {NoStop}%
\end{thebibliography}

%

\end{document}